\documentclass[11pt,a4paper]{article}

\usepackage[T1]{fontenc}
\usepackage[utf8]{inputenc}
\IfFileExists{lmodern.sty}{\usepackage{lmodern}}{}
\usepackage[english]{babel}
\usepackage[left=2.5cm,right=2.5cm,top=2.5cm,bottom=2.5cm]{geometry}
\usepackage{amsmath,amssymb}
\usepackage{graphicx}
\usepackage{booktabs}
\usepackage{rotating}
\usepackage{xcolor}
\usepackage{caption}
\usepackage[numbers,sort&compress]{natbib}
\usepackage[hidelinks]{hyperref}

\graphicspath{{figures/}}
\newcommand{\ohmcm}{\,\Omega\,\mathrm{cm^{2}}}
\newcommand{\Rint}{R_{\mathrm{int}}}
\newcommand{\RSE}{R_{\mathrm{SE}}}
\newcommand{\Rgb}{R_{\mathrm{gb}}}
\newcommand{\Rg}{R_{\mathrm{g}}}
\newcommand{\Rc}{R_{\mathrm{c}}}
\newcommand{\Rct}{R_{\mathrm{ct}}}
\newcommand{\Rcell}{R_{\mathrm{cell}}}
\newcommand{\Rzero}{R_{0}}
\newcommand{\Zcpe}{Z^{\mathrm{CPE}}}
\newcommand{\tauint}{\tau_{\mathrm{int}}}
\newcommand{\taugb}{\tau_{\mathrm{gb}}}
\newcommand{\EA}{E_{\mathrm{A}}}
\newcommand{\kB}{k_{\mathrm{B}}}
\newcommand{\nc}{n_{\mathrm{c}}}
\newcommand{\rc}{r_{\mathrm{c}}}
\newcommand{\NaSmSiO}{Na$_5$SmSi$_4$O$_{12}$}
\newcommand{\NaYSiO}{Na$_5$YSi$_4$O$_{12}$}
\newcommand{\NaSICONc}{Na$_{3.4}$Zr$_2$Si$_{2.4}$P$_{0.6}$O$_{12}$}
\newcommand{\BASE}{Na-$\beta''$-alumina}

\newcommand{\fithead}{%
$T$ [$^\circ$C] & $\Rg$ [$\Omega\,\mathrm{cm}^2$] & $\Rgb$ [$\Omega\,\mathrm{cm}^2$] &
$2\Rint$ [$\Omega\,\mathrm{cm}^2$] & $K_\mathrm{gb}$ [$\mathrm{S\,s^{\alpha}}$] &
$K_\mathrm{int}$ [$\mathrm{S\,s^{\alpha}}$] & $\alpha_\mathrm{gb}$ [-] &
$\alpha_\mathrm{int}$ [-] & $\taugb$ [s] & $\tauint$ [s] \\}

\begin{document}

\begin{center}
{\LARGE\bfseries The Sodium-Metal Interface to Na$_5$RSi$_4$O$_{12}$:\\[2pt]
Stability, Kinetics and Current-Constriction\par}

\vspace{1em}
{\large Ansgar Lowack$^{a,b,*}$, Rafael Anton$^{b}$, Kristian Nikolowski$^{b}$ and Alexander Michaelis$^{a,b}$}
\vspace{0.6em}
{\small\itshape $^{a}$Institute of Material Science, Dresden University of Technology,
Helmholtzstraße 7, 01069 Dresden, Germany\\
\small\itshape $^{b}$Fraunhofer Institute for Ceramic Technologies and Systems (IKTS),
Winterbergstraße 28, 01277 Dresden, Germany\\[3pt]
$^{*}$ Corresponding author: ansgar.lowack@tu-dresden.de}
\end{center}

\vspace{0.8em}

\section*{Abstract}
\noindent
The rare-earth silicates Na$_5$RSi$_4$O$_{12}$ (NaRSiO) reach room-temperature
Na$^+$ conductivities above $2\,\mathrm{mS\,cm^{-1}}$. That makes them promising
solid electrolytes for sodium batteries based on oxide-ceramic ion conductors. The
range of possible cells runs in principle from high-temperature Na--NiCl$_2$ cells
to room-temperature solid-state batteries (SSBs). To assess that prospect we
examine two things, the resistance of the interface to metallic sodium and the
thermodynamic and kinetic stability of the material class against reduction. The
two most relevant chemistries are NaYSiO (R\,=\,Y) and NaSmSiO (R\,=\,Sm). Neither
is thermodynamically stable against sodium metal, which rules them out for
high-temperature application. However, at room temperature the
Na/NaRSiO interface shows no resistive degradation on a timescale of years, so the
decomposition kinetics are slow enough to warrant application in SSBs. At room temperature the
interfacial resistance can be reduced to a level at which no separate interfacial contribution is identifiable in the impedance spectra. Larger values are shown to arise from current constriction rather than
from slow charge-transfer kinetics.

\vspace{0.6em}
\noindent\textbf{Keywords:} sodium solid-state battery, sodium-metal anode,
Na$_5$RSi$_4$O$_{12}$, interfacial resistance, constriction resistance, charge transfer kinetics

\vspace{1em}

\section*{Nomenclature}
\noindent{\small All resistances are area-specific with respect to the geometric
electrode area and are quoted in $\Omega\,\mathrm{cm^{2}}$. $\EA(X)$ denotes the activation energy obtained from an Arrhenius fit
of the quantity $X$.}

\vspace{0.4em}
\noindent{\small
\begin{tabular}{@{}p{1.35cm}@{}p{5.9cm}@{\hspace{0.9em}}p{1.35cm}@{}p{5.9cm}@{}}
$A$ & geometric electrode area & $\Rg$ & grain resistance of the electrolyte \\
$\EA$ & activation energy & $\Rgb$ & grain-boundary resistance of the electrolyte \\
$e$ & elementary charge & $\Rint$ & resistance of a single Na/SE interface \\
$j_0$ & exchange current density & $\RSE$ & effective electrolyte resistance, $\Rg+\Rgb$ \\
$K$ & prefactor of a constant-phase element & $\Rzero$ & pre-exponential factor of an Arrhenius fit \\
$\kB$ & Boltzmann constant & $T$ & absolute temperature \\
$\nc$ & number of contact spots per unit area & $Z$ & complex impedance \\
$\rc$ & radius of a contact spot & $\Zcpe$ & impedance of a constant-phase element \\
$\Rc$ & constriction resistance & $\alpha$ & exponent of a constant-phase element \\
$\Rct$ & charge-transfer resistance & $\sigma$ & ionic conductivity of the electrolyte \\
$\Rcell$ & total resistance of the symmetric cell & $\taugb$, $\tauint$ & relaxation time constants \\
\end{tabular}}

\vspace{0.5em}
\noindent{\small\textbf{Abbreviations:} BASE, \BASE; CLSM, confocal laser scanning
microscopy; CPE, constant-phase element; FESEM, field-emission scanning electron
microscopy; NaRSiO, Na$_5$RSi$_4$O$_{12}$ (R\,=\,rare earth); NaSICON,
Na$_{3.4}$Zr$_2$Si$_{2.4}$P$_{0.6}$O$_{12}$; SE, solid electrolyte; SSB,
solid-state battery; XRD, X-ray diffraction.}

\vspace{1em}

\section{Introduction}
\label{sec:intro}

Replacing the porous, liquid-electrolyte-infiltrated separator of a conventional
cell by a dense solid ion conductor can enable a reversible negative
electrode of pure alkali metal. For sodium, two cell concepts follow from this
substitution.
High-temperature cells operate at $\sim300\,^\circ\mathrm{C}$ with a molten sodium
electrode, a solid-electrolyte (SE) separator and a positive electrode of either molten sulfur or solid NiCl$_2$ infiltrated with a molten NaAlCl$_4$ secondary electrolyte. Room-temperature solid-state batteries (SSBs) combine a solid
sodium electrode, an SE separator and a composite cathode of conventional design.
The merits of either concept are set out elsewhere \citep{sudworth2001, Janek.2023,
Lowack.2025b} and are not repeated here.

This work concerns the rare-earth silicates Na$_5$RSi$_4$O$_{12}$
(R\,=\,rare earth; NaRSiO) \citep{shannon1978} as separators against sodium
metal and addresses two open questions: whether the high interfacial resistances
reported for this class so far are intrinsic to the silicate framework or a
consequence of incomplete contact, and whether NaRSiO is sufficiently stable
against reduction by sodium metal, at high and at room temperature. The
requirements a separator has to meet, and the established materials NaRSiO
competes with, are summarised first.

A wide range of solids, among them sulfides, halides and oxides, conduct Na$^+$
well at room temperature \citep{Hayashi2019, Hu2024, Ma.2020}. However,
conductivity alone does not make a suitable SE separator for rechargeable
batteries. Three further properties decide whether a cell can be cycled reversibly
at useful power density. \emph{(i) Mechanical stability.} On charge, sodium is
plated at the Na/SE interface and may penetrate the separator, something observed
both for solid and molten sodium \citep{richman1975, brennan1980}. To prevent
this, the SE must have a sufficiently high shear modulus to withstand classical
dendrite penetration, which applies only in the case of solid sodium
\citep{monroe2005}. Besides dendrite growth, even stiff SEs are penetrated above a
critical current density if pressure buildup in surface flaws at the Na/SE
interface exceeds their tensile strength \citep{buechele1983degradation,
Lowack.2025, lowack2026analytical}.
\emph{(ii) Low interfacial resistance $\Rint$.} A contemporary single-layer
liquid-electrolyte cell has a total area-specific resistance of $25$--$40\ohmcm$
\citep{Kubal2022}. A single Na/SE interface should stay well inside that budget,
so as not to limit power density and efficiency. That makes
$\Rint\lesssim10\ohmcm$ a sensible orientation value.
\emph{(iii) Chemical stability.} The low resistance has to persist over the life
of the cell. This excludes any SE that sodium reduces to a resistive decomposition
layer. It is the reason why sulfides and halides are problematic for application in direct
contact with sodium metal, despite their high conductivities \citep{wenzel2016,
riegger2021}.

Requirement (i) can be met with oxide-ceramic SEs, as discussed in our previous
work \citep{lowack2026analytical}, which leaves (ii) and (iii) to be assessed.
Three oxide-ceramic families are relevant for that assessment. \BASE\ (BASE) and
NaSICON are the established separators against sodium metal. The rare-earth
silicates NaRSiO are
an interesting third option, because they reach comparable conductivities at
significantly lower sintering temperatures \citep{Anton.2025}. Of the
substitutions compared systematically so far, R\,=\,Yb, Y, Gd and Sm
\citep{Schilm.2022, Anton.2025}, yttrium is the established reference of the
literature, while samarium is technically the more attractive one. It carries the
largest ionic radius of the series, which widens the silicate framework. That
lowers the crystallisation temperature further and should in principle also make
it the best conductor of the series \citep{Schilm.2022}. A clear experimental
confirmation of the conductivity claim is still outstanding, and samarium is the
demanding case for it: its ceramics still contain a large fraction of the
sodium-rich secondary phase Na$_{16.5}$R$_{2.5}$Si$_{12}$O$_{36}$, about
$34\,\mathrm{wt.\%}$ against about $4\,\mathrm{wt.\%}$ for R\,=\,Y
\citep{Schilm.2022}. For a stability test against sodium metal that same secondary
phase makes R\,=\,Sm the most informative choice (Section~\ref{sec:res-stab}).
Both rare earths are among the cheaper ones, roughly an order of magnitude below
neodymium or praseodymium oxide and two to three orders below terbium
\citep{USGS.2026, USGS.2025y}, which matters for a separator needed in bulk.

Requirement (ii) is harder to assess than it appears. Three contributions add in
series and cannot easily be separated from a single impedance spectrum: the
geometric constriction resistance $\Rc$ of an incompletely contacted solid/solid
interface \citep{Holm1967, Eckhardt.2022}, the charge-transfer resistance
$\Rct=\kB T/(e\,j_0)$ as described by the linearised Butler--Volmer equation for a
single-electron transfer (with Boltzmann constant $\kB$, absolute temperature $T$,
elementary charge $e$ and exchange current density $j_0$), and the resistance of
an interphase formed by reduction or surface contamination. Reported values for
chemically similar interfaces scatter accordingly, from several $10^{4}\ohmcm$
down to $\sim1\ohmcm$. Where the interface has been prepared carefully, current
constriction tends to dominate. For Li/LLZO and Li/LPSCl, $\Rint<1\ohmcm$ has been
reported as an upper bound at room temperature \citep{Krauskopf2020, singh2023}.
Higher values on the same materials are traced to a $\Rc$ that contact pressure
suppresses \citep{krauskopf2019toward, zhang2020, Lowack.2025c}. For sodium at
room temperature, Ortmann et al.\ obtained $\Rint\approx2\ohmcm$ on Na/NaSICON
\citep{Ortmann.2023}. On BASE the lowest reported room-temperature values are
$4.6\ohmcm$ against Na--In \citep{boron2025key} and $8\ohmcm$ against sodium
\citep{bay2020}. For NaRSiO, by contrast, the reported values lie far above the
budget set out above, at $364\ohmcm$ \citep{Huttl.2022} and $25\ohmcm$ for
\NaYSiO\ \citep{yang2022}. Whether this is intrinsic to the silicate framework or,
as for the other two families, a contact problem, has been open.

Requirement (iii), in turn, separates the same three families along the
distinction between thermodynamic and kinetic stability. BASE is stable against
sodium at both high and room temperature, as decades of service in
high-temperature cells and first-principles work confirm \citep{sudworth2001,
lacivita.2019}. NaSICON is reduced by molten sodium at $300\,^\circ\mathrm{C}$
\citep{schmid1982}. However, at room temperature the reaction self-passivates,
because its products conduct electrons poorly, and it contributes a technically
irrelevant $\sim1\ohmcm$ \citep{Ortmann.2023}. NaSICON is therefore
thermodynamically unstable but kinetically stable. For NaRSiO neither case is well
established, and this study sets out to answer the stability question against
sodium metal at both temperatures.

\begin{figure}[tbp]
\centering
\includegraphics[width=\linewidth]{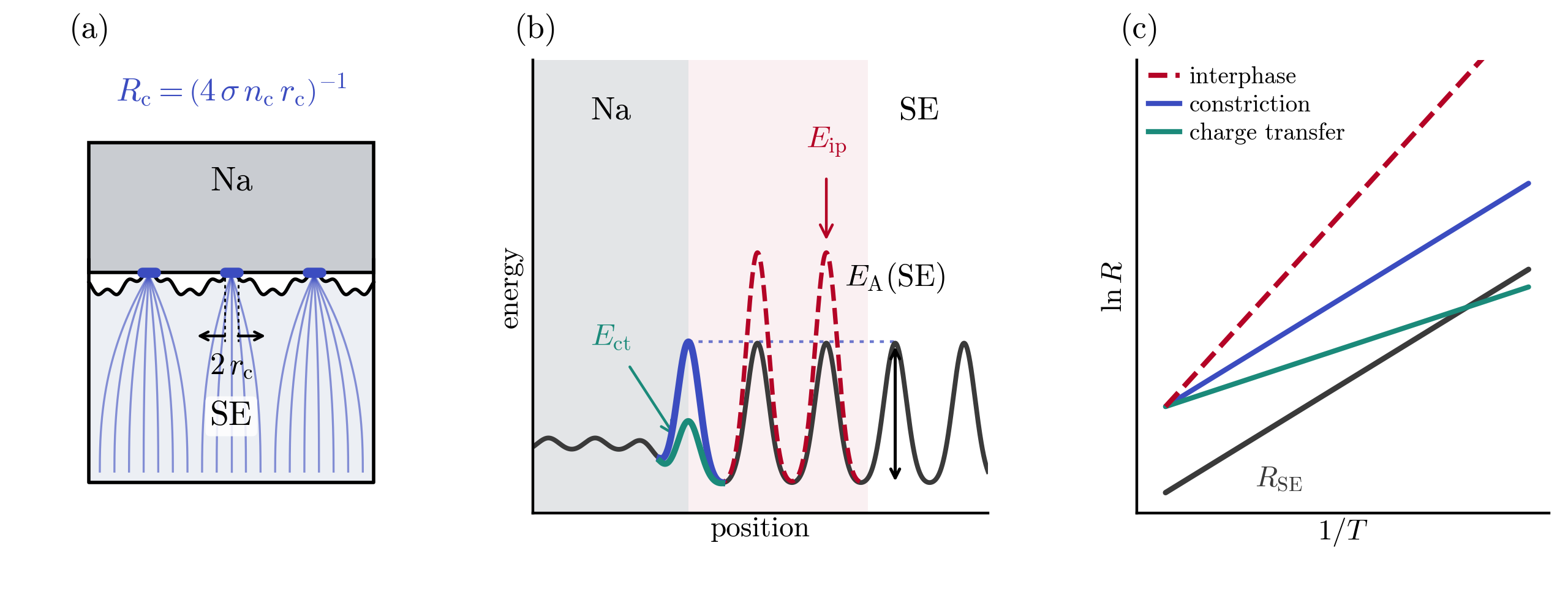}
\caption{Separation of the three contributions to $\Rint$ by
their temperature dependence. (a) Incomplete contact forces the current through $\nc$ contact spots per
unit area, each of radius $\rc$. The resulting constriction resistance contains
only geometry and the electrolyte conductivity $\sigma(T)$, and therefore inherits
$\EA(\RSE)$ exactly. (b) Energy landscape along the transport path. Grey is common
to all three cases. The barrier at the interface is the lattice barrier
$\EA(\mathrm{SE})$ itself when contact alone limits, shown in blue, where the
dotted line marks the equality; the blue barrier is highlighted, not
raised. It is \emph{lower} when a charge-transfer step at
a clean metal/oxide contact limits, shown in green. An interphase of reduction
products instead inserts a layer of \emph{higher} barriers behind the contact,
shown in red. (c) Consequence in the Arrhenius plot. Only the
contact-limited case runs parallel to $\RSE$: constriction raises
the resistance only through the geometric prefactor, which shifts the blue line
upward without changing its slope. Schematic, not to scale.}
\label{fig:scheme}
\end{figure}

We address both open questions, the origin of the interfacial resistance and the stability against sodium metal, on \NaSmSiO, with \NaYSiO\ included in the
stability experiments, using the temperature dependence of the impedance as a
discriminator (Fig.~\ref{fig:scheme}). Only one of the three contributions is
geometric: $\Rc\approx(4\sigma\nc\rc)^{-1}$ \citep{Holm1967} contains the ionic
conductivity $\sigma$ of the electrolyte, the number $\nc$ of contact spots per
unit area and their radius $\rc$. It therefore carries exactly
$\EA(\RSE)$, the activation energy of the effective electrolyte resistance,
whatever that happens to be. The other two carry barriers of their
own. A
charge-transfer-limited interface should show a shallower Arrhenius slope
than the electrolyte and an interphase-limited one a steeper slope (Section~\ref{sec:res-int}), while current
constriction reproduces the electrolyte slope. The three contributions are in
series, so the measured barrier is their resistance-weighted mean. The diagnostic
weight therefore lies in the coincidence of $\EA(\Rint)$ with $\EA(\RSE)$ and not
in the sign of a deviation. We compare a mirror-polished and a deliberately
roughened electrolyte surface on this basis, and then test the interface for
thermodynamic stability (immersion in liquid sodium at $300\,^\circ\mathrm{C}$) and
for kinetic stability ($597$ days of room-temperature storage).

\section{Experimental}
\label{sec:exp}

\textbf{Material.} \NaSmSiO\ powder was prepared by glass crystallisation
\citep{Anton.2025}, uniaxially pressed and sintered in air at
$1050\,^\circ\mathrm{C}$ for $1\,\mathrm{h}$ to pellets of
$\varnothing\,12\,\mathrm{mm}\times1.3\,\mathrm{mm}$ with a relative density
$>97\,\%$ ($\sim99\,\%$ by image analysis of the micrographs). Besides the main
phase, the known and likewise conductive Na$_{16.5}$Sm$_{2.5}$Si$_{12}$O$_{36}$ is
present \citep{Anton.2025}. Diffractograms and micrographs are given in the
Supporting Information (Section~\ref{sec:phase}). Blocking-electrode measurements on Cu$|$SE$|$Cu cells
serve here as the reference for the \emph{barrier} of ion transport inside the SE
and give $\EA(\sigma T)=0.28\,\mathrm{eV}$, in agreement with previous
observations \citep{Schilm.2022, Anton.2025}. This value follows from an Arrhenius
fit of $\sigma T$, which is the convention for a random-walk hopping conductivity;
the resistances below are fitted without that prefactor, and the two conventions
differ by $\kB T\approx0.025\,\mathrm{eV}$ (Section~\ref{sec:blocking} of the Supporting Information). The spectra and the Arrhenius analysis are given there as well. The
absolute conductivity obtained from them,
$\sigma(30\,^\circ\mathrm{C})\approx2\,\mathrm{mS\,cm^{-1}}$, is treated as a
lower bound: On one and the same pellet, sputtered copper or gold electrodes
consistently give a somewhat higher resistance than pressed sodium. We ascribe
this to incomplete wetting of the sputtered film. The symmetric Na$|$SE$|$Na cells
of this work imply slightly higher conductivities, and only the activation energy,
not the magnitude, is taken from the blocking measurement.
\NaYSiO, used only for the immersion experiment, was prepared identically but
sintered at $1120\,^\circ\mathrm{C}$.

\textbf{Interfaces.} All preparation took place in an argon glovebox. This is
essential here, because a few minutes of exposure to air at $\sim30\,\%$ relative
humidity measurably degrade the SE surface. As a precaution, all SE samples were therefore annealed at
$700\,^\circ\mathrm{C}$ for $4\,\mathrm{h}$ in argon prior to cell assembly, which
reverses such damage by removing adsorbed water and contamination (Supporting
Information, Section~\ref{sec:humidity}). SE samples were ground from both
sides, plane-parallel in a custom holder, either with P360 SiC paper
(\emph{rough}, approximately reproducing the as-sintered grain-scale topography)
or stepwise to P4000 (\emph{polished}, mirror finish).
Grinding under argon removes any surface contamination formed during sintering in
air, so that any non-geometric contribution to $\Rint$ cannot be a preparation
artefact. It also removes a substantial amount of material, so that the separator
of the assembled cells is appreciably thinner than the as-sintered pellet,
$0.75\pm0.01\,\mathrm{mm}$. This is deliberate and increases the
sensitivity of the interfacial resistance measurement, because the thinner the
separator, the larger the share of the total resistance that the interfaces carry.
Confocal laser scanning microscopy (CLSM) on a reference oxide gives relative
surface-area factors (true area over geometric cross-section) of $1.75\pm0.27$
as-sintered, $1.45\pm0.06$ after P360 and $1.05\pm0.02$ after P4000. Comparable
magnitudes are assumed here (Supporting Information, Section~\ref{sec:topography}). Sodium ($\geq99.0\,\%$) was
rolled to $\sim500\,\mathrm{\mu m}$, scraped clean with a steel microblade
immediately before use, applied to the freshly prepared SE surfaces by hand, and
the resulting stacks were pressed at $97\,^\circ\mathrm{C}$ and
$10\,\mathrm{MPa}$ for about one minute. Cell stacks were subsequently
sealed in PAT-Core-Force housings (EL-CELL) at $\approx10\,\mathrm{MPa}$ and held
at $30\,^\circ\mathrm{C}$ under load for at least two hours before measurement.

\begin{figure}[tbp]
\centering
\includegraphics[width=0.4\linewidth]{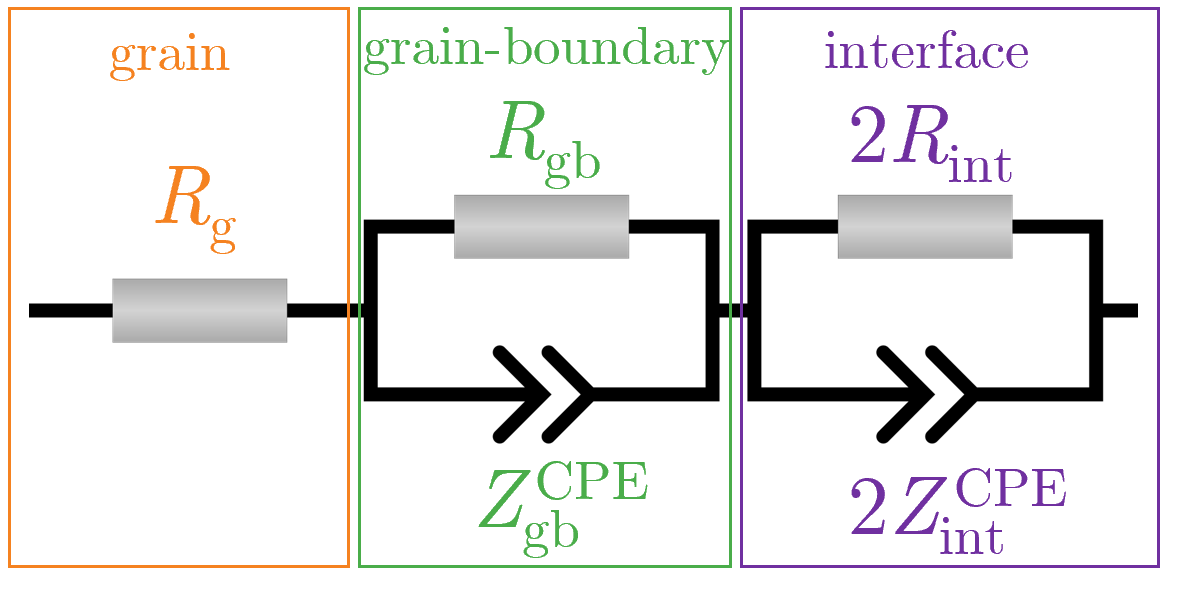}
\caption{Equivalent circuit used to fit the impedance spectra of the symmetric
Na$|$SE$|$Na cells. The grain contribution enters as a pure resistance $\Rg$,
because the grain capacitance is not resolvable below $1\,\mathrm{MHz}$. The
grain-boundary and the interfacial contribution are each a resistance in parallel
with a constant-phase element. The two Na/SE interfaces of a symmetric cell are in
series, so the fitted interfacial element is $2\Rint$ and $\Rint$ per interface is
half of it.}
\label{fig:circuit}
\end{figure}

\textbf{Impedance spectroscopy.} Impedance was measured using a four-wire setup on
a VMP3 (BioLogic) in a temperature chamber (BINDER) ($\Delta T=\pm1\,^\circ\mathrm{C}$). The
measurements were potentiostatic with $10\,\mathrm{mV}$ amplitude, from
$1\,\mathrm{MHz}$ to $1\,\mathrm{Hz}$, ten points per decade, five averaged runs,
and ran from $-30$ to $30\,^\circ\mathrm{C}$ in $10\,\mathrm{K}$ steps. Spectra
were fitted with RelaxIS (rhd instruments) to the equivalent circuit of
Fig.~\ref{fig:circuit},
$\Rg+(\Rgb\!\parallel\!\Zcpe_{\mathrm{gb}})
+(2\Rint\!\parallel\!2\Zcpe_{\mathrm{int}})$, in which $\Rg$ and $\Rgb$ are the
grain and the grain-boundary resistance of the electrolyte and $\Zcpe$ is the
impedance of a constant-phase element (CPE). Because the two Na/SE interfaces of a
symmetric cell are in series, the fitted interfacial element is $2\Rint$, and
$\Rint$ per interface is half of it. Time constants are
$\tau=(R\,K)^{1/\alpha}$, evaluated from the \emph{absolute} resistance of the
element (in $\Omega$) and from the CPE prefactor $K$ and exponent $\alpha$
belonging to it. Data above
$0.7\,\mathrm{MHz}$ (cable inductance,
$>20\,\%$ error) and below the onset of anodic interfacial dynamics
($\sim10\,\mathrm{Hz}$ at $-30\,^\circ\mathrm{C}$, rising to
$\sim300\,\mathrm{Hz}$ at $30\,^\circ\mathrm{C}$ \citep{Ortmann.2023,
Eckhardt.2022}) were excluded. The grain capacitance is not resolvable below
$1\,\mathrm{MHz}$, so $\Rg$ enters as a pure resistance and the
effective electrolyte resistance $\RSE=\Rg+\Rgb$ is evaluated. All resistances are
area-specific with respect to the geometric electrode area $A$, which was
$0.50\,\mathrm{cm^{2}}$ for the polished \NaSmSiO\ cell and
$1.13\,\mathrm{cm^{2}}$ for all other cells, and $\Rint$ is
given per interface. Resistances were fitted to
$R(T)=\Rzero\exp(\EA/\kB T)$ with the pre-exponential factor $\Rzero$.
Uncertainties follow from Gaussian
propagation, and the $\EA$ errors include $\Delta T=\pm1\,^\circ\mathrm{C}$. All
individual spectra and the complete set of fit parameters are given in Section~\ref{sec:nasmsio-spectra}
of the Supporting Information. Cells with NaSICON as SE material
(Na$_{3.4}$Zr$_2$Si$_{2.4}$P$_{0.6}$O$_{12}$,
sintered at $1350\,^\circ\mathrm{C}$ for $30\,\mathrm{min}$) were prepared in a
similar fashion and measured with the
identical protocol. They serve as a validation of the method presented here
against the one Na/oxide interface for which an independent literature value
exists \citep{Ortmann.2023}, and are documented in Section~\ref{sec:nasicon} of the Supporting Information.

\textbf{Stability tests.} \NaSmSiO\ and \NaYSiO\ pellets were immersed in liquid
sodium at $300\,^\circ\mathrm{C}$ for four weeks, in an MgO crucible inside a
muffle furnace inside an argon-filled glovebox. After that exposure, samples were
freed of adhering sodium mechanically, and imaged in the glovebox at the surface
and on a fresh fracture cross-section. Diffractograms were taken after exposure.
NaSICON pellets were immersed under the same conditions as a reference, and a
further NaSICON pellet for one week at $500\,^\circ\mathrm{C}$ (Section~\ref{sec:nasicon-immersion} of the
Supporting Information). For
the ageing experiment we used a symmetric cell with rough interfaces from an
earlier batch. Its separator is thicker, $1.8\,\mathrm{mm}$. Its absolute
resistance is therefore about three times higher than that of the cells of
Section~\ref{sec:res-int}. The cell was
first measured $\sim2\,\mathrm{h}$ after contacting, stored in argon at room
temperature under $3$--$10\,\mathrm{MPa}$, and re-measured at
$30\,^\circ\mathrm{C}$ after $4$, $30$, $90$, $415$ and $597$ days.

\section{Results and discussion}

\subsection{The interfacial resistance is contact-limited}
\label{sec:res-int}

\begin{figure}[tbp]
\centering
\includegraphics[width=0.78\linewidth]{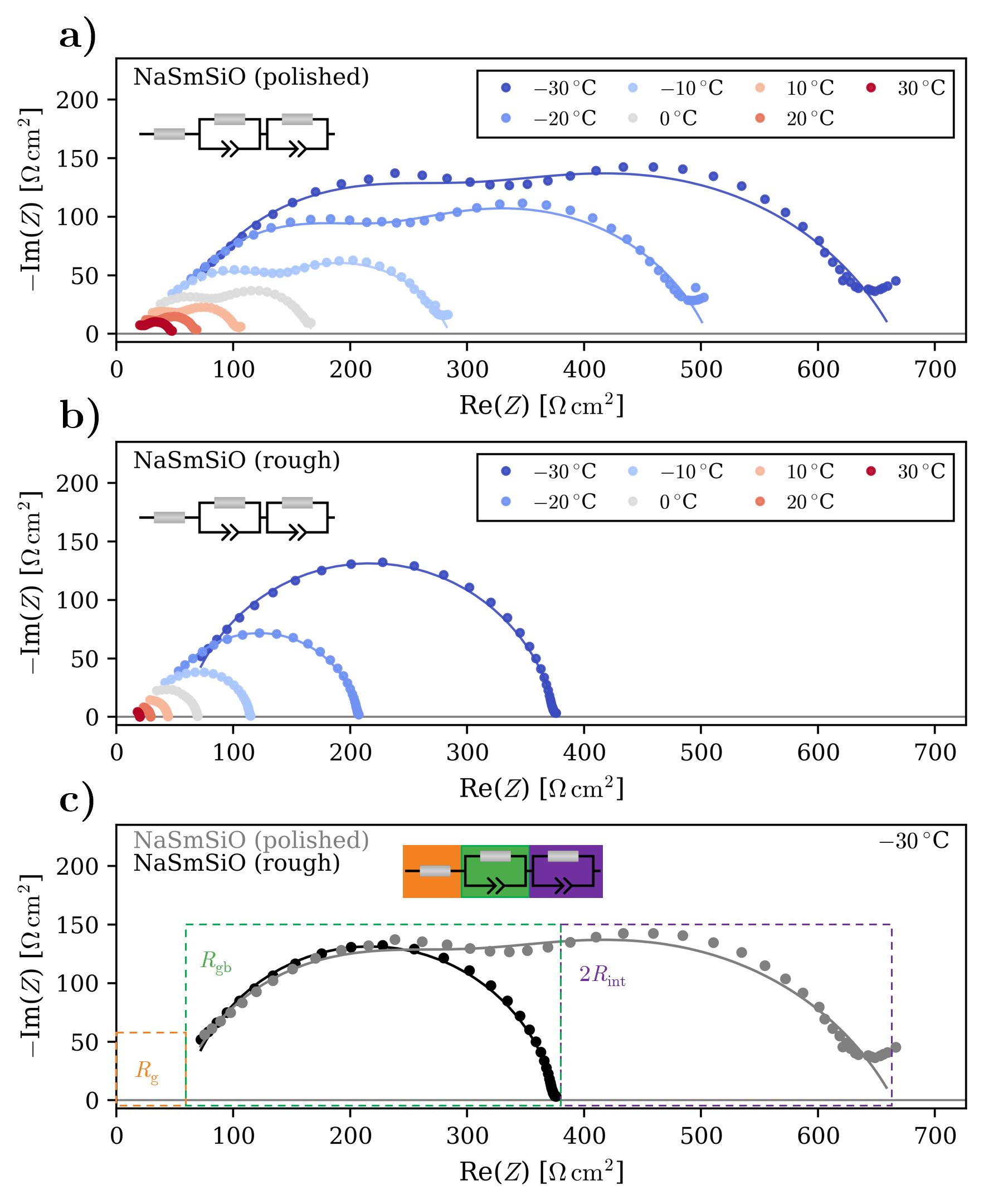}
\caption{Nyquist plots of symmetric Na$|$\NaSmSiO$|$Na cells: (a) polished and
(b) rough electrolyte surfaces between $-30$ and $30\,^\circ\mathrm{C}$, and
(c) the two $-30\,^\circ\mathrm{C}$ spectra superimposed with the assignment of the
equivalent-circuit elements. Roughening removes the interfacial arc. Symbols:
data; lines: fits.}
\label{fig:impedance}
\end{figure}

The deviations from $\EA(\RSE)$ to be expected from charge transfer and from an
interphase have opposite signs (Fig.~\ref{fig:scheme}b,c). Charge
transfer across a clean metal/oxide contact should be \emph{faster} than bulk
migration rather than slower: no solvation shell has to be formed or stripped, the
sodium leaves a shallow metallic site instead of a deep lattice site, and the metal
screens the charge of the ion as it forms. Atomistic calculations for Li/LLZO put
the transfer step at $\approx0.09\,\mathrm{eV}$ against $\approx0.25\,\mathrm{eV}$
for the bulk hop \citep{Iwasaki.2024}, consistent with the fast transfer kinetics
measured on that interface \citep{Krauskopf2020}. An interphase of reduction
products behaves the other way round. It conducts worse than the electrolyte it
replaces, because otherwise it would not raise the resistance. The sodium oxides
and simple silicates formed in such reactions typically carry higher migration
barriers, so here a steeper slope is expected.

The polished cell shows two separated polarisation processes, resolved most clearly
at $-30\,^\circ\mathrm{C}$, where the thermally activated transport shifts
resistances up and characteristic frequencies down (Fig.~\ref{fig:impedance}a,c).
The high-frequency arc is the grain-boundary polarisation, that is, the
low-frequency tail of $\RSE$. The grain polarisation itself is unresolved. The
lower-frequency arc is $\Rint$. The time constants support the assignment:
$\tauint$ exceeds $\taugb$ by a factor of $22$ to $32$ throughout
(Table~\ref{tab:results}). Furthermore, $\EA(\tauint)=0.28\pm0.05\,\mathrm{eV}$
agrees with $\EA(\Rint)$, as expected, because the capacitive contributions are
essentially temperature-independent. A significant discrepancy would indicate that
the second arc had been misassigned. Above $10\,^\circ\mathrm{C}$ the
grain-boundary polarisation moves close to the high-frequency limit of the setup,
which inflates the uncertainty of its fit parameters; $\Rg$ and $\Rgb$ become
strongly anticorrelated there. Their individual errors grow to several tens of per cent,
while their sum $\RSE$ stays well determined, because the sum is what the arc
actually fixes.

\begin{table}[tbp]
\centering
\caption{Fit results for Na$|$\NaSmSiO$|$Na. Time constants are given as the range
from $-30$ to $30\,^\circ\mathrm{C}$. $\Rint$ is area-specific per interface. For
the rough cell the interfacial arc is not separable in the one-dimensional model,
so only the bound $\Rint\lesssim\RSE/20$ at $30\,^\circ\mathrm{C}$ is quoted.}
\label{tab:results}
\small
\begin{tabular}{lcc}
\toprule
 & \textbf{polished} & \textbf{rough} \\
\midrule
$\EA(\RSE)$ (eV) & $0.28\pm0.02$ & $0.31\pm0.01$ \\
$\EA(\Rint)$ (eV) & $0.30\pm0.01$ & not separable \\
$\Rint(30\,^\circ\mathrm{C})$ ($\Omega\,\mathrm{cm^{2}}$) & $13.6\pm0.9$ & $\lesssim1$ \\
$\taugb$ (s) & $(1.7\pm2.5)\cdot10^{-6}$ -- $(1.1\pm6.0)\cdot10^{-7}$
             & $(2.81\pm0.24)\cdot10^{-6}$ -- $(9.8\pm1.5)\cdot10^{-8}$ \\
$\tauint$ (s) & $(3.8\pm2.0)\cdot10^{-5}$ -- $(3.6\pm1.4)\cdot10^{-6}$ & -- \\
\bottomrule
\end{tabular}
\end{table}

\begin{figure}[tbp]
\centering
\includegraphics[width=0.80\linewidth]{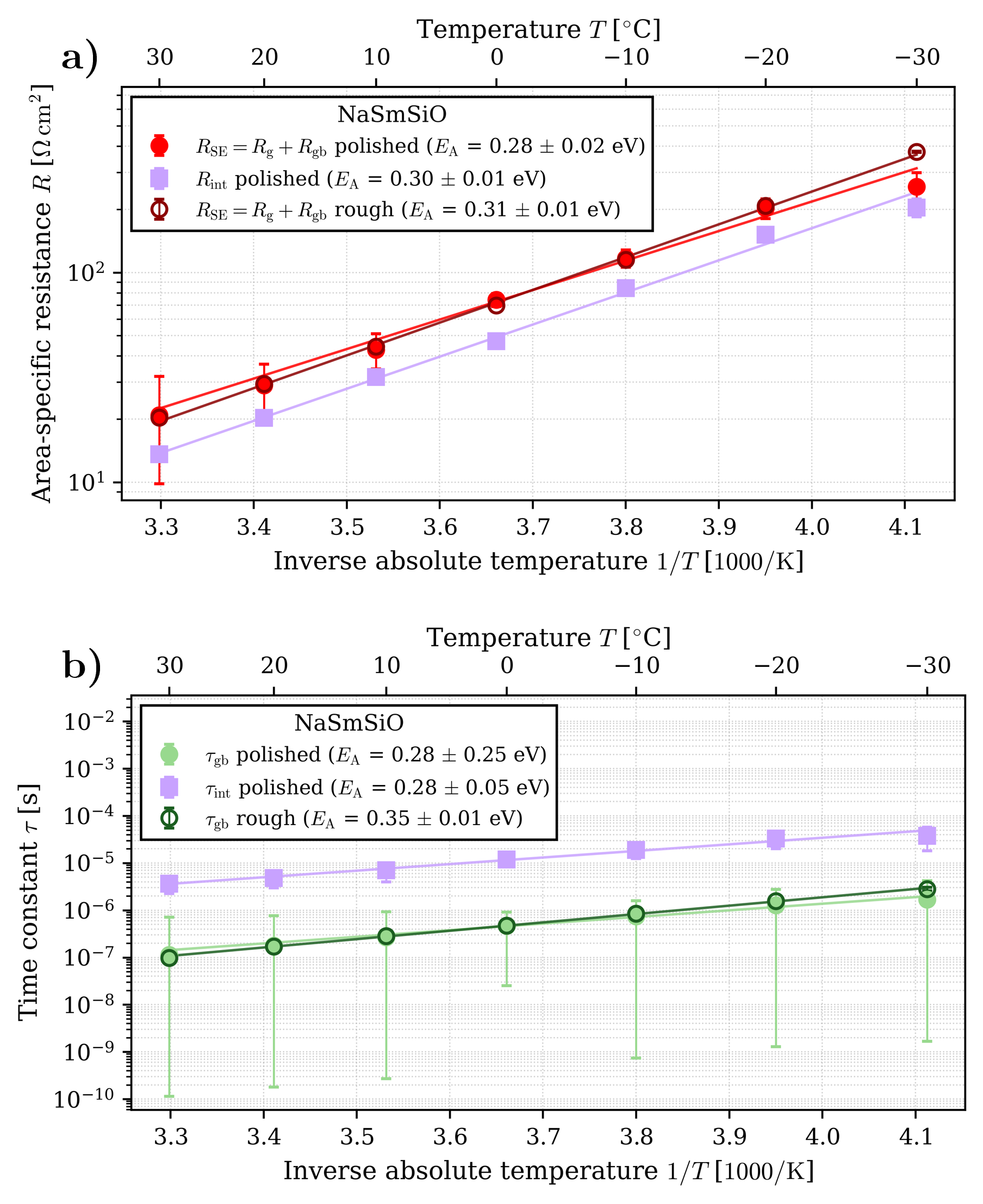}
\caption{Arrhenius plots for Na$|$\NaSmSiO$|$Na between $-30$ and
$30\,^\circ\mathrm{C}$: (a) effective electrolyte resistance $\RSE$ for both
surface finishes and interfacial resistance $\Rint$ of the polished cell;
(b) the corresponding time constants $\taugb$ and $\tauint$. Fits to
$R(T)=\Rzero\exp(\EA/\kB T)$.}
\label{fig:arrhenius}
\end{figure}

On the polished surface $\Rint$ reaches the magnitude of $\RSE$ itself, so the
interface would dominate a thin-separator cell. Its activation energy is
$0.30\pm0.01\,\mathrm{eV}$ and that of the electrolyte resistance is
$0.28\pm0.02\,\mathrm{eV}$. The two differ by no more than the measurement
accuracy. The electrolyte value is in turn consistent with the bulk
ion-conduction barrier of this material measured on blocking electrodes, after
converting between the $\sigma T$ and the resistance fitting convention
(Section~\ref{sec:blocking}).
By the criterion of Section~\ref{sec:intro} the constriction resistance
is therefore the dominant contribution: charge transfer or an interphase would
impose a barrier of unrelated magnitude and a visibly different slope in
Fig.~\ref{fig:arrhenius}a. Notably, no interphase signature appears even though
\NaSmSiO\ is strongly moisture-sensitive, which confirms that grinding under argon
leaves the surface clean enough.

Roughening confirms the diagnosis directly. A single arc then remains even at
$-30\,^\circ\mathrm{C}$ (Fig.~\ref{fig:impedance}b), so the equivalent circuit becomes
overdetermined; $\Rint$ was fixed to zero \emph{as a model parameter} and the whole
spectrum attributed to $\RSE$. That this reading is justified follows
from two independent observations. First, the spectra are qualitatively identical
to the polished ones except for the missing interfacial arc
(Fig.~\ref{fig:impedance}c). Second, the remaining arc stays at the time constant
of the grain-boundary process of the polished cell, with an activation energy
compatible with it within the combined uncertainty (Table~\ref{tab:results}).
The interfacial process is more than an
order of magnitude slower, and it is simply absent. Only an upper
bound can therefore be given. Requiring $2\Rint$ to lie at least one order of
magnitude below $\RSE$, i.e.\ $\Rint\lesssim\RSE/20$, and inserting the measured
$\RSE(30\,^\circ\mathrm{C})=20.4\pm1.0\ohmcm$ yields
$\Rint\lesssim1\ohmcm$.

That the bound is meaningful, and not an artefact of fixing $\Rint$ to zero, is
confirmed on a reference system. Na$|$NaSICON$|$Na cells prepared and evaluated
in exactly the same way behave identically. They are constriction-dominated when
polished and show no separable interfacial arc when roughened. They give
$\Rint\lesssim1.5\ohmcm$ at $30\,^\circ\mathrm{C}$. That is consistent with the
$\Rint\approx2\ohmcm$ Ortmann et al.\ obtained for the same material by a
different characterisation route, namely high-resolution impedance spectroscopy
and degradation measurements on differently assembled cells \citep{Ortmann.2023}.
Because our cells are hot-pressed and then equilibrated for hours before the
measurement, the self-passivating decomposition layer resolved there is already
contained in that number. The agreement is thus a cross-check of two independent
methods (Supporting Information, Section~\ref{sec:nasicon}).

This places Na/\NaSmSiO\ at the level established for Li/LLZO, Li/LPSCl and
Na/NaSICON \citep{Krauskopf2020, singh2023, Ortmann.2023}. To our knowledge it is
the first NaRSiO interface reported there. It lies at least an order of magnitude
below the $364$ and $25\ohmcm$ previously published for this material class
\citep{Huttl.2022, yang2022}. The high values reported previously are thus not
intrinsic to the
silicate framework but a consequence of surface preparation and contacting. The
gain is also not an area effect. The true surface area increases only from $1.05$
to $1.45$ times the geometric cross-section, a factor of $1.4$ against more than an
order of magnitude in resistance. Improved wetting and mechanical interlocking of
the plastically deforming metal on the rougher substrate are the plausible origin,
which the present data do not resolve further.

One caveat matters for everything that follows under load. A vanishing $\Rc$ does
not imply a homogeneously contacted interface. With Holm's parallel-patch
approximation $\Rc\approx(4\sigma\nc\rc)^{-1}$
\citep{Holm1967}, $\Rc\sim1\ohmcm$ at $1\,\mathrm{mS\,cm^{-1}}$ requires only
$\nc\rc=250\,\mathrm{cm^{-1}}$. That is met for instance by
$1\,\mathrm{\mu m}$ patches at $2.5\cdot10^{6}\,\mathrm{cm^{-2}}$, a contacted
area fraction of $7.9\,\%$. The interfaces reported here are therefore in theory compatible
with $\sim90\,\%$ of the area being out of contact, and with local current
densities an order of magnitude above the nominal value. That is irrelevant for the
stationary state measured here, but it is the relevant quantity for pore growth on
stripping \citep{Fuchs.2024, lowack2026experimental} and for metal penetration on plating
\citep{richman1975, Lowack.2025c, lowack2026analytical}.

\subsection{Kinetically, not thermodynamically, stable}
\label{sec:res-stab}

A low initial resistance is only useful if it survives the service life of a cell.
For the two established SE families this is settled (Section~\ref{sec:intro}). For
NaRSiO, both the thermodynamic and the kinetic stability against sodium metal remained unresolved.

\begin{figure}[tbp]
\centering
\includegraphics[width=0.56\linewidth]{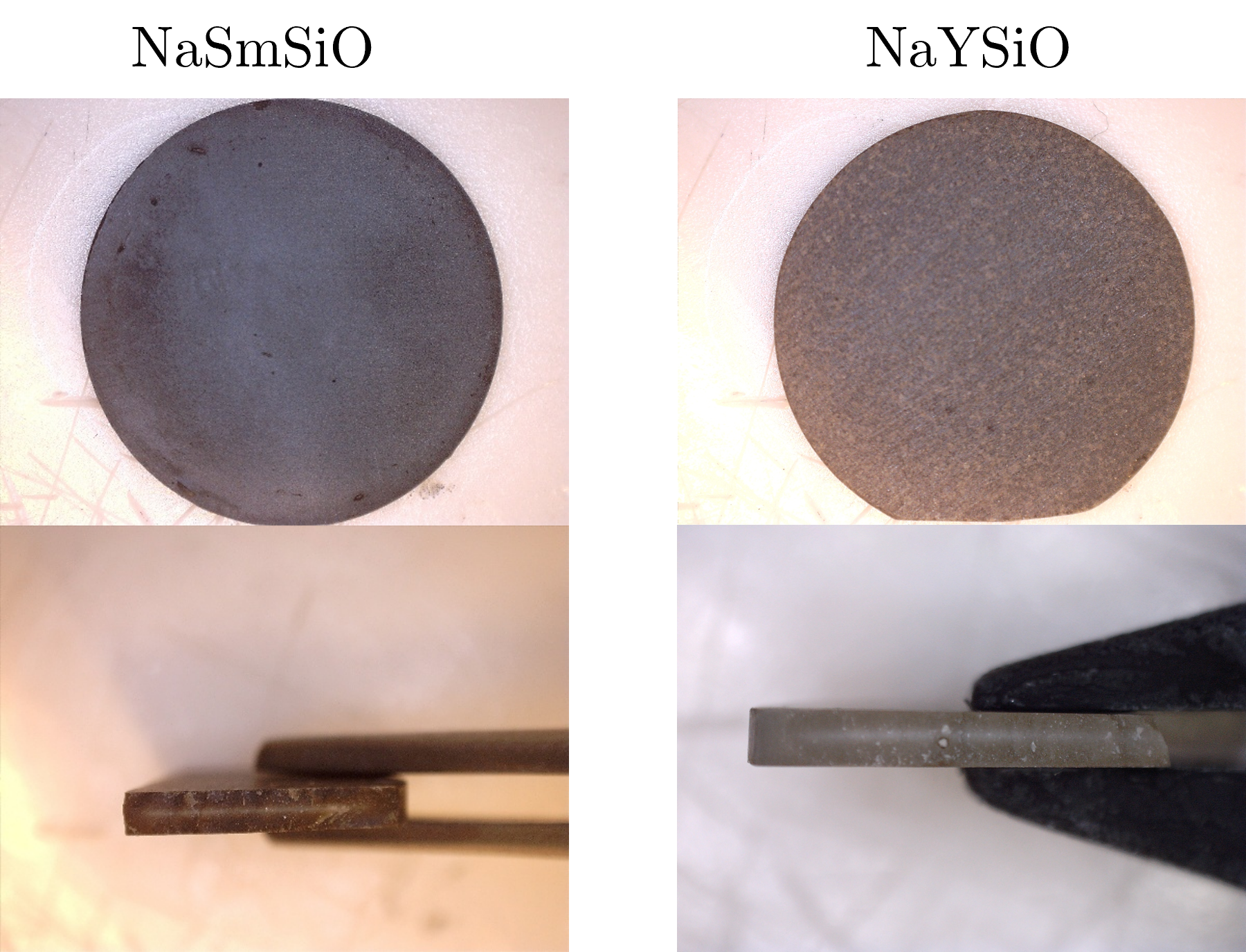}
\caption{\NaSmSiO\ (left) and \NaYSiO\ (right) pellets
($\varnothing\sim12\,\mathrm{mm}$, $\sim1.3\,\mathrm{mm}$) after four
weeks in liquid
sodium at $300\,^\circ\mathrm{C}$. The originally white ceramics blacken (top) and
the fracture cross-sections (bottom) show that the discolouration penetrates the
pellet. The reaction does not passivate.}
\label{fig:immersion}
\end{figure}

After four weeks in liquid sodium at $300\,^\circ\mathrm{C}$, \NaSmSiO\ and
\NaYSiO\ pellets are dark grey throughout their cross-section, with only a thin
white core left (Fig.~\ref{fig:immersion}). The discolouration appears on the fresh
fracture surface, which excludes SiC debris from the cleaning step. It persists
after prolonged water contact, which excludes residual metallic sodium. Both
materials are therefore thermodynamically unstable against sodium metal. At
$300\,^\circ\mathrm{C}$, the reaction front does not self-passivate. The
comparison is direct: NaSICON
pellets immersed under identical conditions discolour only at the surface and keep
a white core, and they need $500\,^\circ\mathrm{C}$ to react through
(Supporting Information, Section~\ref{sec:nasicon-immersion}). Against molten sodium, NaRSiO and NaSICON
are therefore both unstable, but the decomposition of NaSICON is slower and
self-limiting. X-ray diffraction of the exposed pellets
remains inconclusive and shows essentially the starting phases. Crystalline
secondary phases are detected only from $\sim1$--$5\,\mathrm{wt.\%}$, whereas a few
tenths of a percent of a strongly absorbing species already suffice to blacken an
oxide ceramic, and the expected products need not be crystalline. Comparing the two
host lattices still constrains the mechanism. Sm$^{3+}\!\rightarrow$Sm$^{2+}$
reduction would explain the colour for \NaSmSiO, but yttrium has no divalent state
accessible under these conditions and \NaYSiO\ discolours very similarly. Reduction
of the rare-earth cation can therefore not be the sole cause. Reduction of the
silicate framework to elemental silicon is the most plausible remaining route.
NaRSiO, like NaSICON, is consequently unsuitable for high-temperature sodium
batteries and cannot replace BASE there.

Room-temperature use is not excluded by this, for kinetic reasons. With a plausible
$0.5$--$1\,\mathrm{eV}$ for such a reaction, the Arrhenius rate ratio between $300$
and $30\,^\circ\mathrm{C}$ is $8.2\cdot10^{3}$ to $6.8\cdot10^{7}$. That turns
weeks into centuries or even millions of years. As expected, Na/NaRSiO interfaces
stored at room temperature show no discolouration after more than a year. However,
optical inspection is not sufficient to prove that decomposition is irrelevant at
room temperature, since an interphase on the $\sim1\,\mathrm{nm}$ scale of the
sodium hopping distance would already drive up $\Rint$. The interfacial resistance
is therefore investigated again by impedance spectroscopy.

Within the NaRSiO family, \NaSmSiO\ is the conservative material for this test.
It crystallises with the highest fraction of the secondary phase
Na$_{16.5}$Sm$_{2.5}$Si$_{12}$O$_{36}$ of the substitutions studied so far,
$\sim34\,\mathrm{wt.\%}$ against $\sim4\,\mathrm{wt.\%}$ for \NaYSiO\
\citep{Schilm.2022, Anton.2025}. As this secondary phase might be less stable
against sodium than the main phase, \NaSmSiO\ puts an upper rather than a lower
bound on the degradation to be expected from the material class.

\begin{figure}[tbp]
\centering
\includegraphics[width=\linewidth]{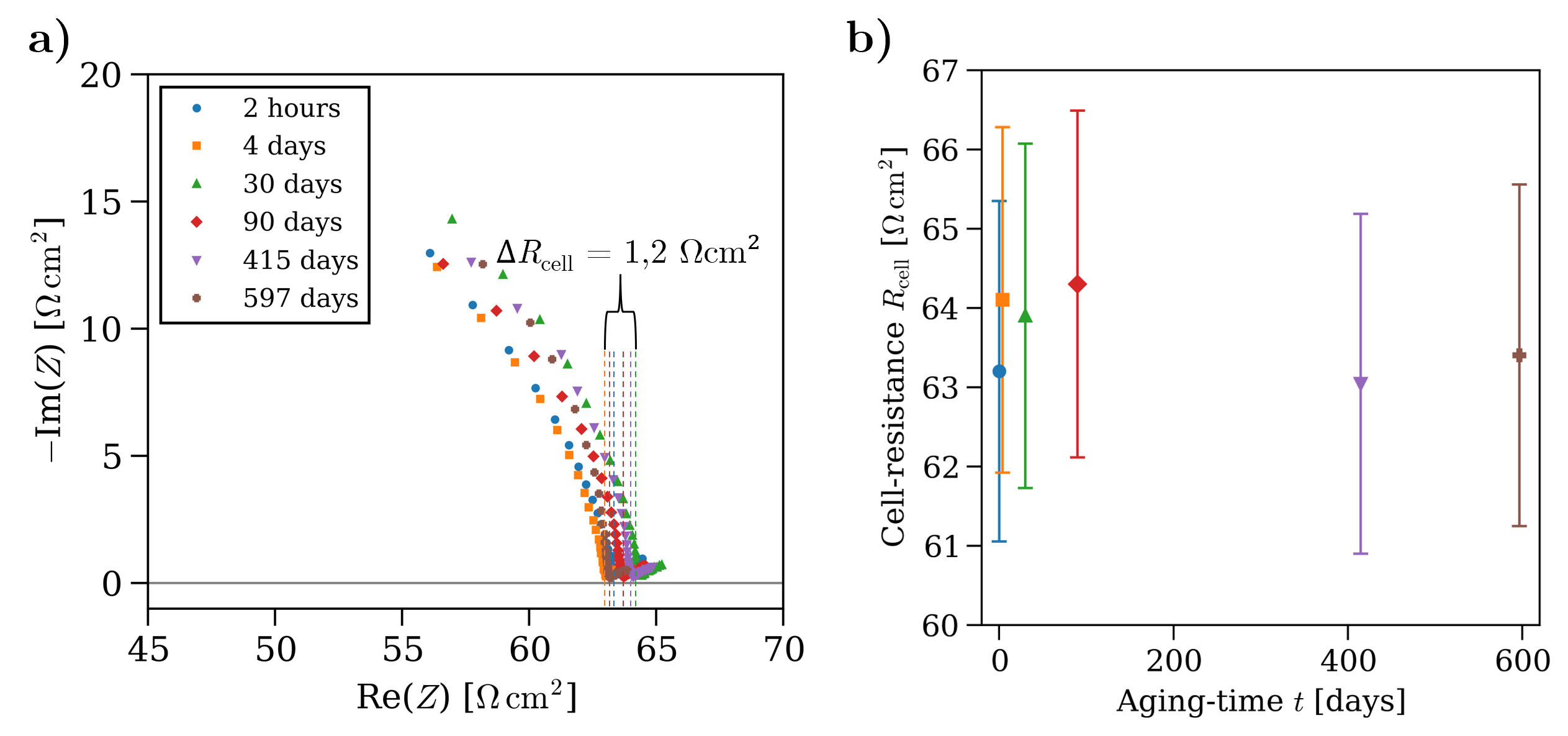}
\caption{Room-temperature ageing of a symmetric Na$|$\NaSmSiO$|$Na cell with rough
interfaces and a $1.8\,\mathrm{mm}$ separator, from an earlier batch than the cells
of Fig.~\ref{fig:impedance}. (a) Nyquist plots at $30\,^\circ\mathrm{C}$ after the
indicated storage times. The low-frequency minimum of $|\mathrm{Im}(Z)|$ gives the
cell resistance $\Rcell=\Rg+\Rgb+2\Rint$, which shifts by at most $1.2\ohmcm$.
(b) $\Rcell$ versus ageing time, with the error bars implied by
$\Delta T=\pm1\,^\circ\mathrm{C}$. No significant ageing is measurable.}
\label{fig:ageing}
\end{figure}

Impedance spectra recorded during storage resolve changes of the cell
resistance down to a few per cent. Over $597$ days of
room-temperature
storage the spectra remain qualitatively unchanged, the interfacial resistance
staying buried in the grain-boundary arc (Fig.~\ref{fig:ageing}a). With
$\EA\approx0.3\,\mathrm{eV}$ and $\Delta T=\pm1\,^\circ\mathrm{C}$, the dominant
uncertainty of the total cell resistance $\Rcell=\Rg+\Rgb+2\Rint$ measured at
$30\,^\circ\mathrm{C}$ is
$\Delta\Rcell/\Rcell=(\EA/\kB T^{2})\,\Delta T\approx3.8\,\%$. Within that accuracy
$\Rcell$ does not change, and neither does $\Rint$. The total drift stays below
$1.2\ohmcm$, which is below $2\,\%$ (Fig.~\ref{fig:ageing}b). The Na/\NaSmSiO\
interface is thus kinetically stable on the timescale of years, and the very low
initial interfacial resistance is retained.

\section{Conclusions}

We investigated the interfacial resistance and the stability of
Na$_5$RSi$_4$O$_{12}$ solid electrolytes (R\,=\,Sm, Y) in contact with sodium
metal by temperature-dependent impedance spectroscopy, immersion in liquid sodium
at $300\,^\circ\mathrm{C}$ and room-temperature storage of symmetric cells.
The resistance of the Na/\NaSmSiO\ interface is limited by contact
geometry, not by interfacial kinetics or decomposition chemistry: $\Rint$ is
Arrhenius-like with an activation energy of $0.30\pm0.01\,\mathrm{eV}$, which
matches that of the electrolyte within the measurement accuracy. On a polished
surface the constriction resistance leaves $\Rint$ at the magnitude of the
electrolyte resistance, even at $10\,\mathrm{MPa}$ contact pressure. Sodium pressed
onto a clean surface roughened to the scale of the ceramic grain structure removes
the interfacial arc from the spectra altogether. This gives $\Rint\lesssim1\ohmcm$
at $30\,^\circ\mathrm{C}$, at least an order of magnitude below previous reports
for NaRSiO and on a par with the best alkali-metal/oxide interfaces known. The high
resistances published for this material class are therefore an artefact of surface
preparation rather than a property of the silicate framework.

\NaSmSiO\ and \NaYSiO\ are thermodynamically unstable against sodium metal, with a
non-passivating reaction front penetrating a $\sim1.3\,\mathrm{mm}$ pellet
within four
weeks at $300\,^\circ\mathrm{C}$. This rules the class of SEs out for
high-temperature
sodium batteries. However, at room temperature the interface is kinetically stable over
$597$ days, so the low interfacial resistance is retained on application-relevant
timescales. NaRSiO thus joins NaSICON and BASE as a viable separator against
sodium metal at room temperature. What the impedance data cannot exclude is
pronounced local current focusing: a vanishing constriction resistance is
compatible with a contacted area fraction of only $\sim10\,\%$, which is the
quantity that governs the stability of the interface under load.

\section*{Supporting Information}
The Supporting Information is appended to this article.
Sections~\ref{sec:phase} and~\ref{sec:topography} document the sintered material: XRD and FESEM of \NaSmSiO, and
the CLSM topography of the as-sintered, P360 and P4000 surfaces. Section~\ref{sec:blocking} gives
the ionic conductivity from Cu$|$SE$|$Cu blocking-electrode cells, together with
the Arrhenius analysis. Section~\ref{sec:nasmsio-spectra} collects all individual Nyquist plots of the
Na$|$\NaSmSiO$|$Na cells and the complete equivalent-circuit fit parameters, and
Section~\ref{sec:humidity} the impedance of such cells after deliberate humid-air exposure.
Section~\ref{sec:nasicon} validates the method on Na$|$NaSICON$|$Na cells against the literature
value of Ortmann et al.\ \citep{Ortmann.2023}, with the individual spectra and fit
parameters of those cells. Section~\ref{sec:nasicon-immersion} covers the immersion of NaSICON in liquid
sodium, as a reference for the stability experiment of
Section~\ref{sec:res-stab}.

\section*{Declaration of competing interest}
The authors declare no competing financial interests.

\section*{Data availability}
The data are included in the manuscript and the Supporting Information as
diagrams, and are available as a Zenodo archive
(10.5281/zenodo.24372353).

\section*{Author contributions}
The article is a condensed version of Chapter~4 of A. Lowack's dissertation ``Dynamische Grenzflächenstabilität und
Risswachstum in oxidkeramischen Natriummetall-Feststoffakkumulatoren'', which is
supervised by K. Nikolowski and A. Michaelis. R. Anton synthesized the NaRSiO-powders and evaluated the XRD-analysis.

\section*{Acknowledgements}
The authors thank Qianli Ma and Frank Tietz for providing the reference NaSICON
powder. The authors gratefully acknowledge financial support from the Free State
of Saxony within the M.ERA-NET 2023 project ``Keramisches Anodenmaterial für
definierte Natriumabscheidung'' (Na-CerAnode, grants 100728944 and 100727996) and the European Union within the Horizon Europe project ``Development of Operando Techniques and Multiscale Modeling to Face the Zero-Excess Solid-State Battery Challenge'' (OPERA, grant 101103834).

\section*{Declaration of generative AI and AI-assisted technologies}
During the preparation of this work the authors used Claude Opus~5 (Anthropic,
October 2026) to condense and translate the manuscript into English from
a chapter of A. Lowack's German-language doctoral dissertation. All experimental
work, the equivalent-circuit fits and the interpretation are the authors' own. No
data were generated, measured, analysed or altered by the model. After using this
tool the authors reviewed and edited the content as needed and take full
responsibility for the content of the publication. The model was further used to
help draft the schematic of Fig.~\ref{fig:scheme} in Python.

\clearpage
\appendix
\setcounter{section}{0}
\setcounter{figure}{0}
\setcounter{table}{0}
\setcounter{equation}{0}
\renewcommand{\thesection}{S\arabic{section}}
\renewcommand{\theHsection}{S\arabic{section}}
\counterwithin{figure}{section}
\counterwithin{table}{section}
\renewcommand{\thefigure}{\thesection.\arabic{figure}}
\renewcommand{\thetable}{\thesection.\arabic{table}}
\renewcommand{\theHfigure}{\thefigure}
\renewcommand{\theHtable}{\thetable}

\begin{center}
{\Large\bfseries Supporting Information\par}
\vspace{0.4em}
{\large The Sodium-Metal Interface to Na$_5$RSi$_4$O$_{12}$:
Stability, Kinetics and Current-Constriction\par}
\end{center}
\addcontentsline{toc}{section}{Supporting Information}

\vspace{1em}

\section{Phase composition and microstructure of the sintered \texorpdfstring{\NaSmSiO}{Na5SmSi4O12}}
\label{sec:phase}

\begin{figure}[htb]
\centering
\includegraphics[width=0.92\linewidth]{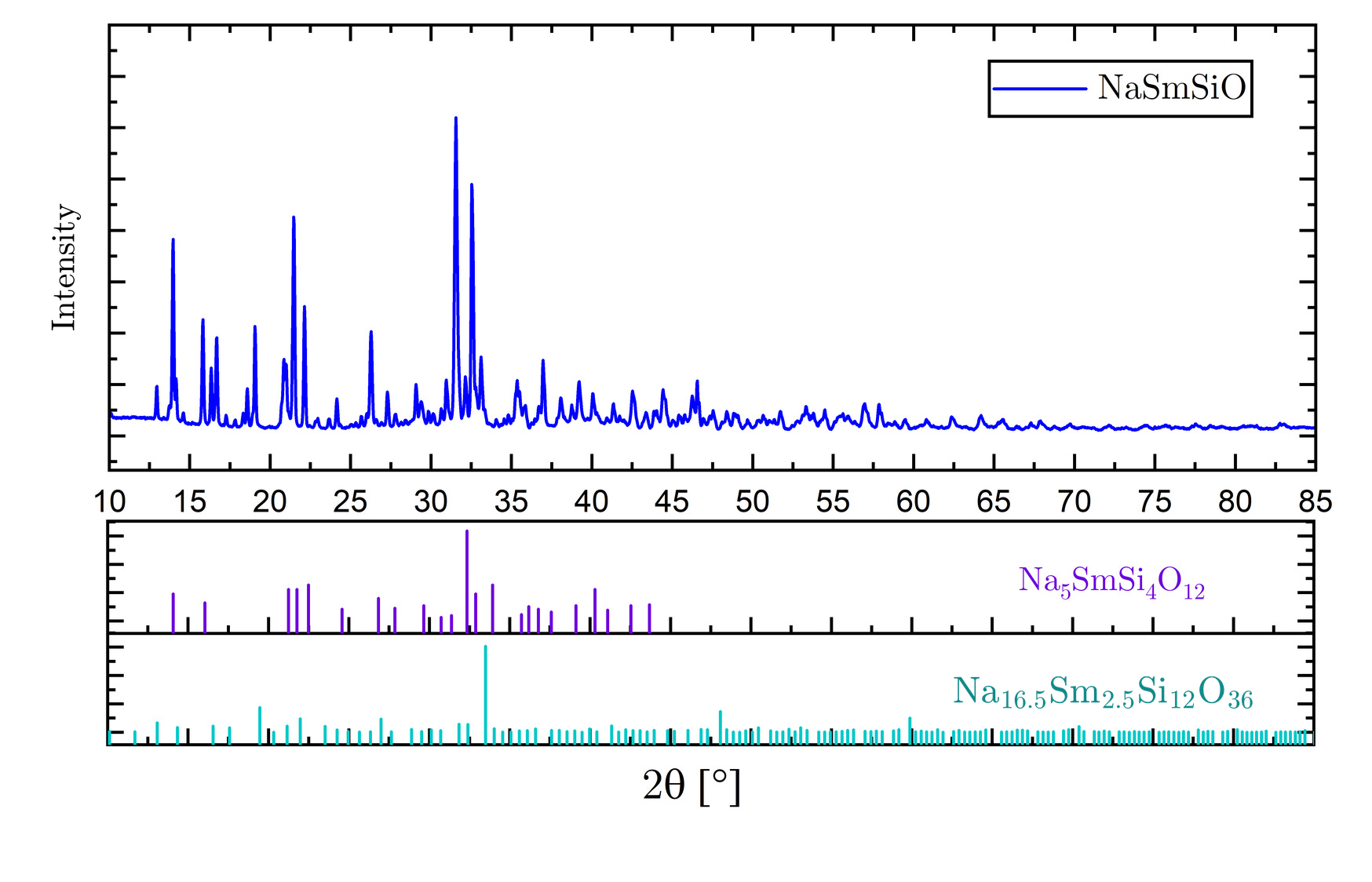}
\caption{X-ray diffractogram of the \NaSmSiO\ ceramic after sintering at
$1050\,^\circ\mathrm{C}$, with the reference patterns of the main phase
Na$_5$SmSi$_4$O$_{12}$ and of the secondary phase
Na$_{16.5}$Sm$_{2.5}$Si$_{12}$O$_{36}$.}
\label{fig:xrd}
\end{figure}

\begin{figure}[htb]
\centering
\includegraphics[width=\linewidth]{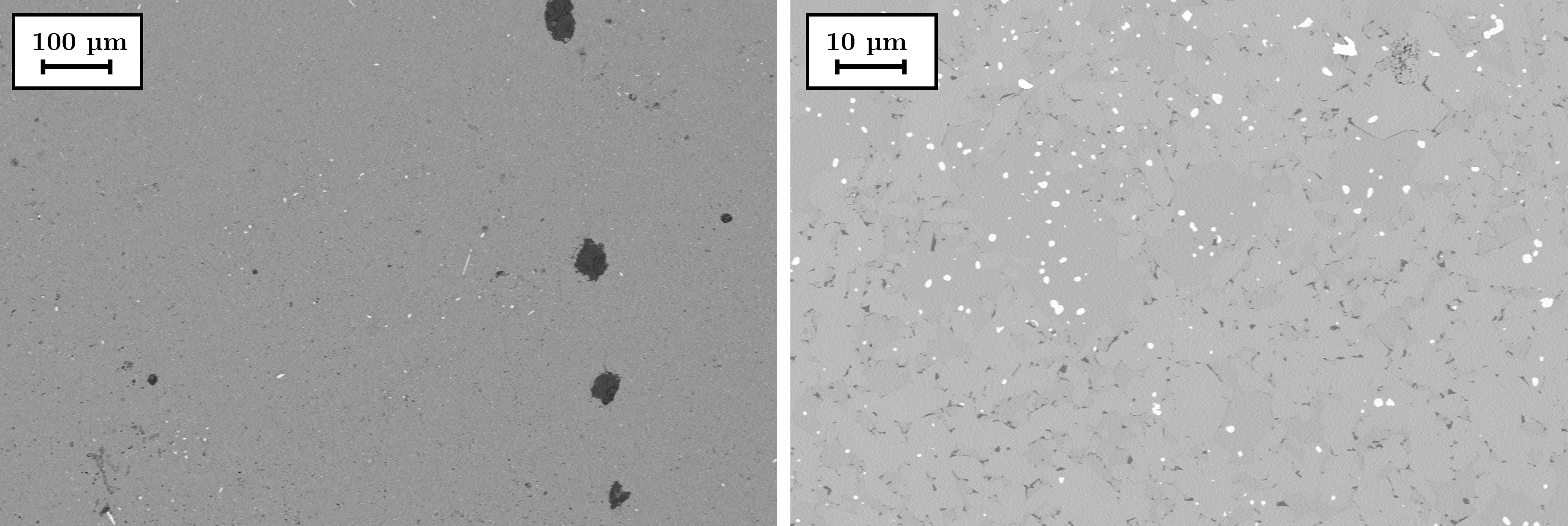}
\caption{FESEM micrographs of the \NaSmSiO\ microstructure at two magnifications.}
\label{fig:fesem}
\end{figure}

Fig.~\ref{fig:xrd} shows the X-ray diffractogram of the sintered pellet. Besides
the \NaSmSiO\ main phase, the sodium-richer
Na$_{16.5}$Sm$_{2.5}$Si$_{12}$O$_{36}$ is present. This phase is known for the
glass-crystallisation route, it is itself a sodium-ion conductor, and it
corresponds to the expected phase coexistence under these conditions
\citep{Anton.2025}.
Fig.~\ref{fig:fesem} shows the microstructure at two magnifications: a dense
sintered body with isolated, closed porosity and minor secondary precipitates at
the grain boundaries. Archimedean weighing gives a relative density above
$97\,\%$, and image analysis of the micrographs gives $\sim99\,\%$. Both are
consistent with a closed pore network.

\section{Surface topography}
\label{sec:topography}

\begin{figure}[htb]
\centering
\includegraphics[width=0.92\linewidth]{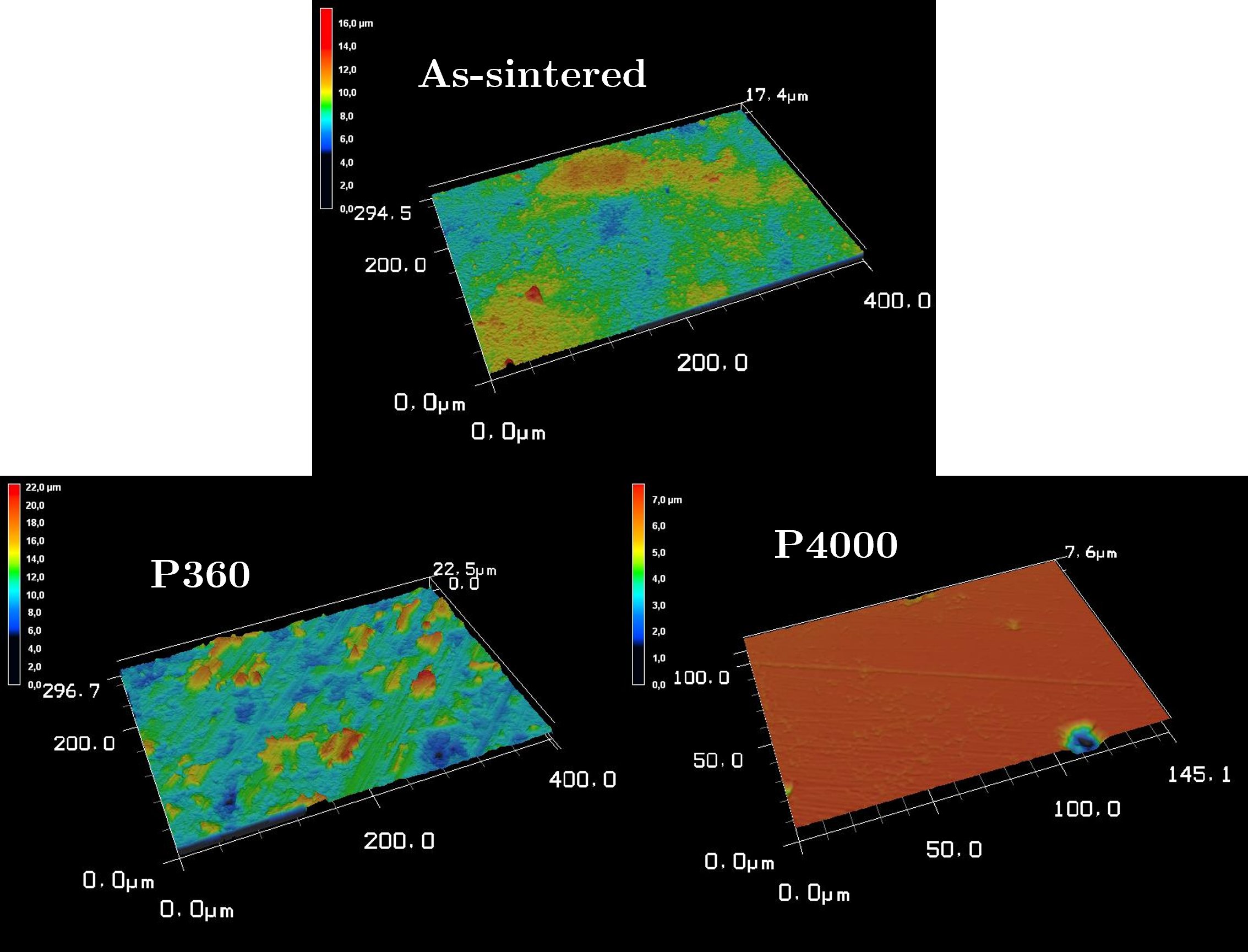}
\caption{Confocal laser scanning micrographs of the as-sintered, P360-ground and
P4000-polished ceramic surface. Height scales as given by the instrument.}
\label{fig:clsm}
\end{figure}

Fig.~\ref{fig:clsm} compares the three surface finishes by confocal laser
scanning microscopy (CLSM). The relative surface-area factors (true three-dimensional
area divided by the geometric cross-section) are $1.75\pm0.27$ for the
as-sintered surface, $1.45\pm0.06$ after grinding with P360 SiC paper and
$1.05\pm0.02$ after polishing to P4000. They were determined on NaSICON reference
samples, and comparable magnitudes are assumed for \NaSmSiO. Note that the optical
resolution limit of the method does not capture the atomic-scale roughness that is
electrochemically decisive. These factors therefore compare preparation states
rather than measure the electrochemically active area.

\section{Ionic conductivity from blocking-electrode measurements}
\label{sec:blocking}

\begin{figure}[htb]
\centering
\includegraphics[width=0.8\linewidth]{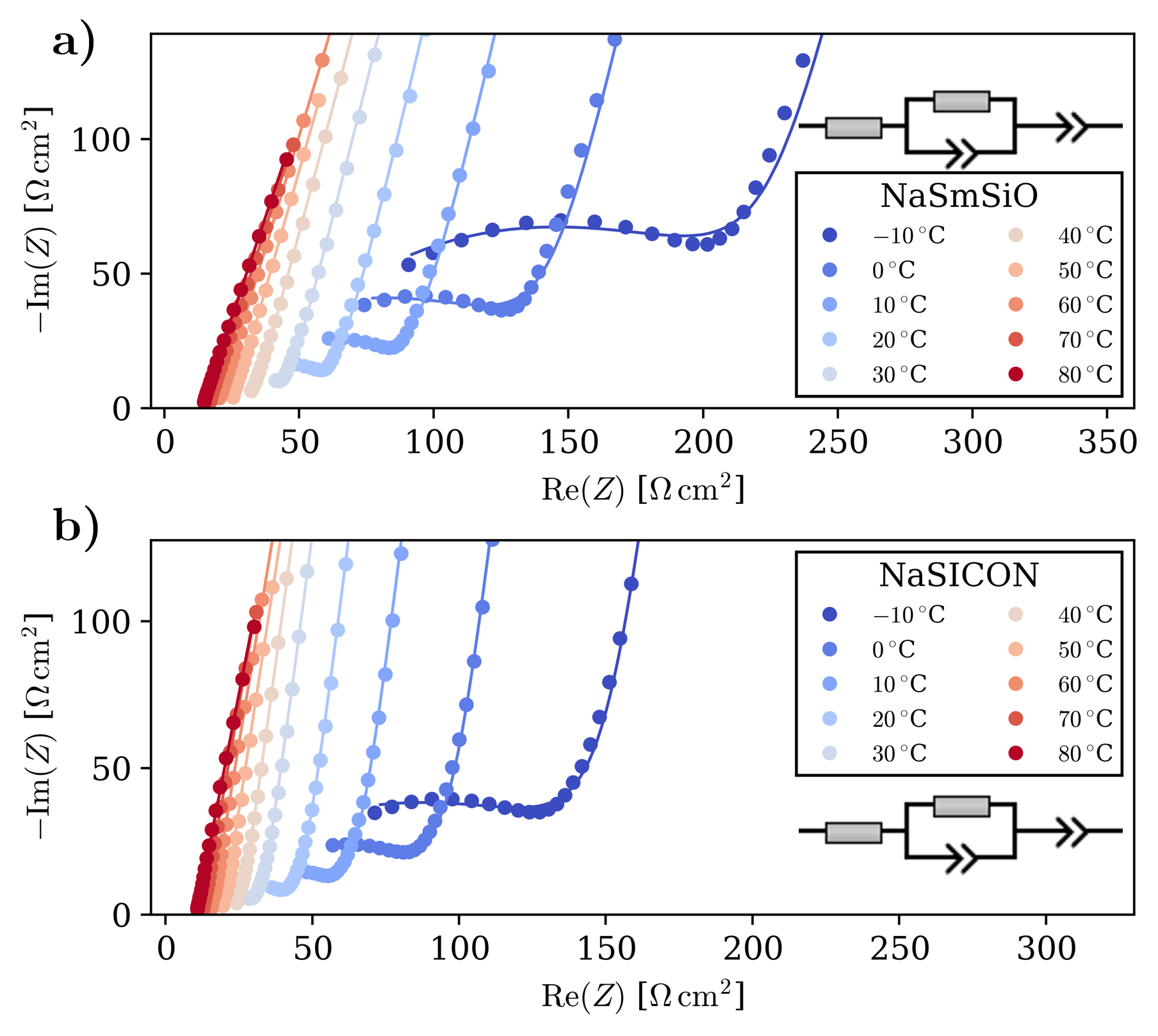}
\caption{Nyquist plots of the impedance spectra of (a) \NaSmSiO\ and (b)
NaSICON with sputtered, ion-blocking copper electrodes between $-10$ and
$+80\,^\circ\mathrm{C}$. Insets: the equivalent circuit used for the fit.
Symbols: data; lines: fits.}
\label{fig:blocking}
\end{figure}

\begin{figure}[htb]
\centering
\includegraphics[width=0.8\linewidth]{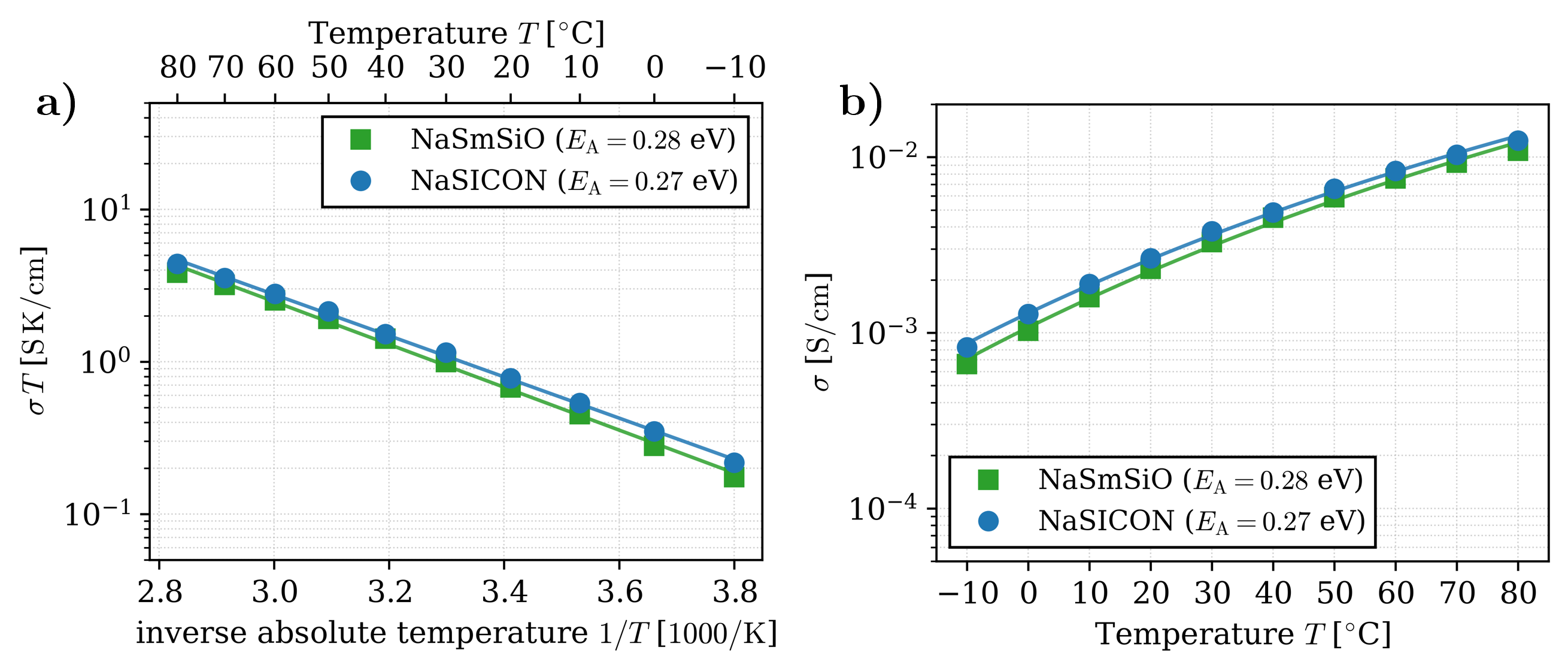}
\caption{Arrhenius analysis of the blocking-electrode data of
Fig.~\ref{fig:blocking}: (a) $\sigma T$ against $1/T$, with the activation energy
from the fit, and (b) the specific conductivity $\sigma$ against temperature.}
\label{fig:blocking-arr}
\end{figure}

The activation energy of bulk sodium-ion transport quoted in the main text was
determined on symmetric cells with sputtered, ion-blocking copper electrodes
between $-10$ and $+80\,^\circ\mathrm{C}$. Fig.~\ref{fig:blocking} shows the corresponding Nyquist
plots together with the equivalent circuit used for the fit.
Fig.~\ref{fig:blocking-arr} shows the resulting Arrhenius analysis. For
\NaSmSiO\ this gives $\EA(\sigma T)=0.28\,\mathrm{eV}$ and
$\sigma(30\,^\circ\mathrm{C})\approx2\,\mathrm{mS\,cm^{-1}}$. For NaSICON it gives
$\EA(\sigma T)=0.27\,\mathrm{eV}$ at a comparable conductivity.

The fitted quantity here is $\sigma T=\sigma_0\exp(-\EA/\kB T)$, as is conventional
for a random-walk hopping conductivity, with the pre-exponential factor
$\sigma_0$. The symmetric-cell resistances of the main text are instead fitted as
$R=\Rzero\exp(\EA/\kB T)$, without the factor $T$. Because
$R\propto1/\sigma=(T/\sigma_0)\exp(\EA/\kB T)$, a straight line through $\ln R$ versus
$1/T$ returns $\EA(R)=\EA(\sigma T)-\kB T_{\mathrm{eff}}$. Here
$T_{\mathrm{eff}}$ is the effective mean temperature of the fitted range,
$304\,\mathrm{K}$ in this section and $271\,\mathrm{K}$ for the symmetric cells.
The values
above therefore correspond to $\EA(R)=0.25\,\mathrm{eV}$ (\NaSmSiO) and
$0.24\,\mathrm{eV}$ (NaSICON). The difference, $\kB T\approx0.025\,\mathrm{eV}$,
is of the same size as the activation-energy differences discussed in the
main text, so the two numbers must not be compared without this conversion. The
criterion of the main text itself is unaffected: $\EA(\Rint)$ and $\EA(\RSE)$
are both obtained from resistances over the same temperature range, so the
prefactor cancels exactly between them.

The absolute conductivity from these cells is treated as a lower bound rather than
as a measured value. Across our samples, cells with sputtered copper or gold
electrodes consistently show a somewhat higher resistance than cells made from the
same pellets, of the same thickness, with pressed sodium electrodes. We have no
quantitative explanation and attribute it to incomplete wetting of the sputtered
film, which adds a contact contribution that the evaluation assigns to the
electrolyte. The effect is systematic and of the order of a few tens of per cent:
the symmetric cells of the main text correspond to
$\sigma(30\,^\circ\mathrm{C})\approx3\,\mathrm{mS\,cm^{-1}}$ against the
$\approx2\,\mathrm{mS\,cm^{-1}}$ measured here. The activation energy is unaffected
by this, because a temperature-independent series contribution changes the
prefactor and not the slope. That is why the blocking-electrode measurement is
used here for $\EA$ only.

\section{Impedance spectra and fit parameters: Na\texorpdfstring{$|$}{|}\texorpdfstring{\NaSmSiO}{Na5SmSi4O12}\texorpdfstring{$|$}{|}Na}
\label{sec:nasmsio-spectra}

\begin{figure}[htb]
\centering
\includegraphics[width=\linewidth]{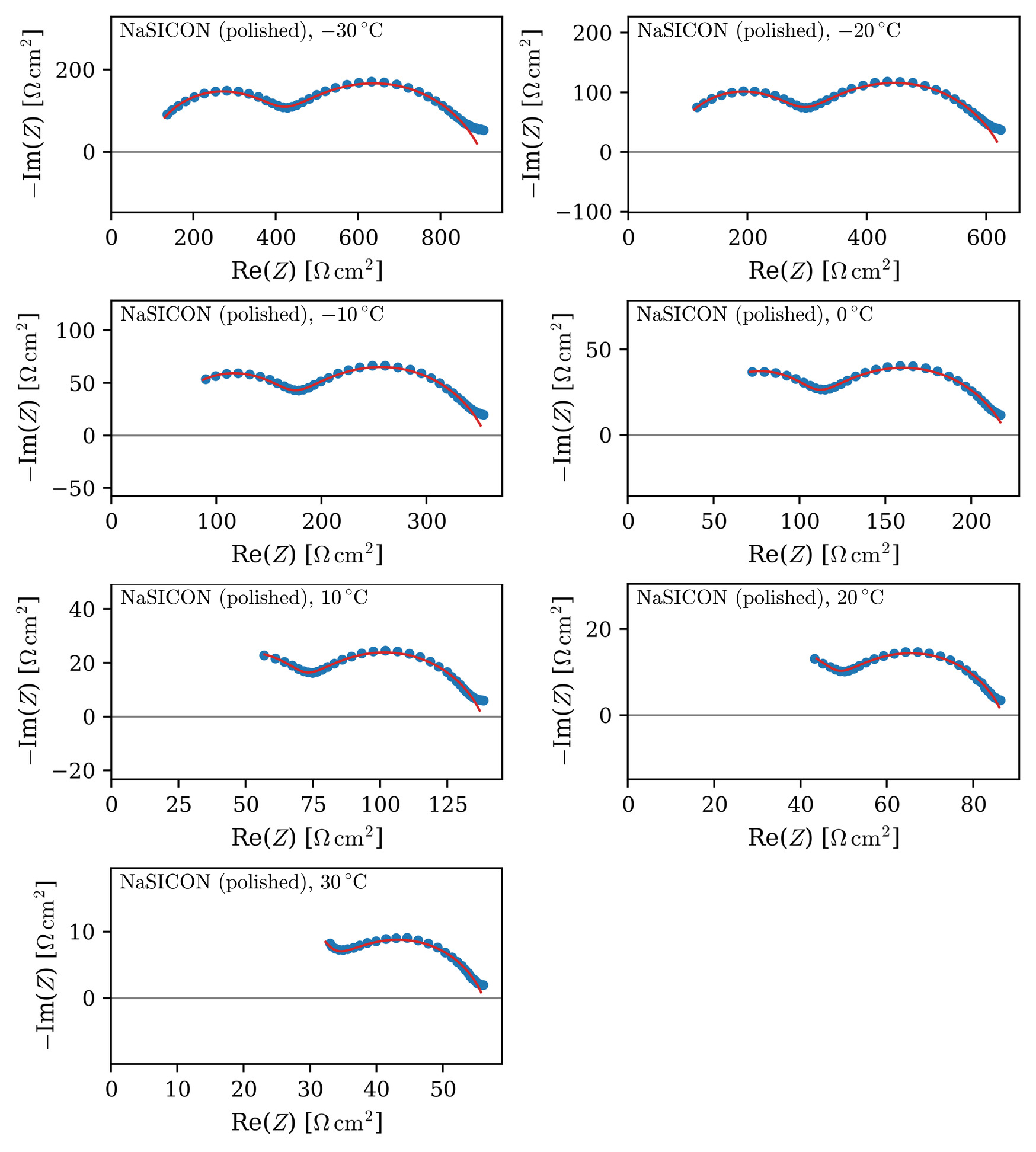}
\caption{Nyquist plots of the individual impedance spectra of symmetric
Na$|$\NaSmSiO$|$Na cells with a \emph{polished} electrolyte surface between
$-30$ and $+30\,^\circ\mathrm{C}$. Symbols: data; lines: fits to the equivalent
circuit.}
\label{fig:nyq-nasmsio-pol}
\end{figure}

\begin{figure}[htb]
\centering
\includegraphics[width=\linewidth]{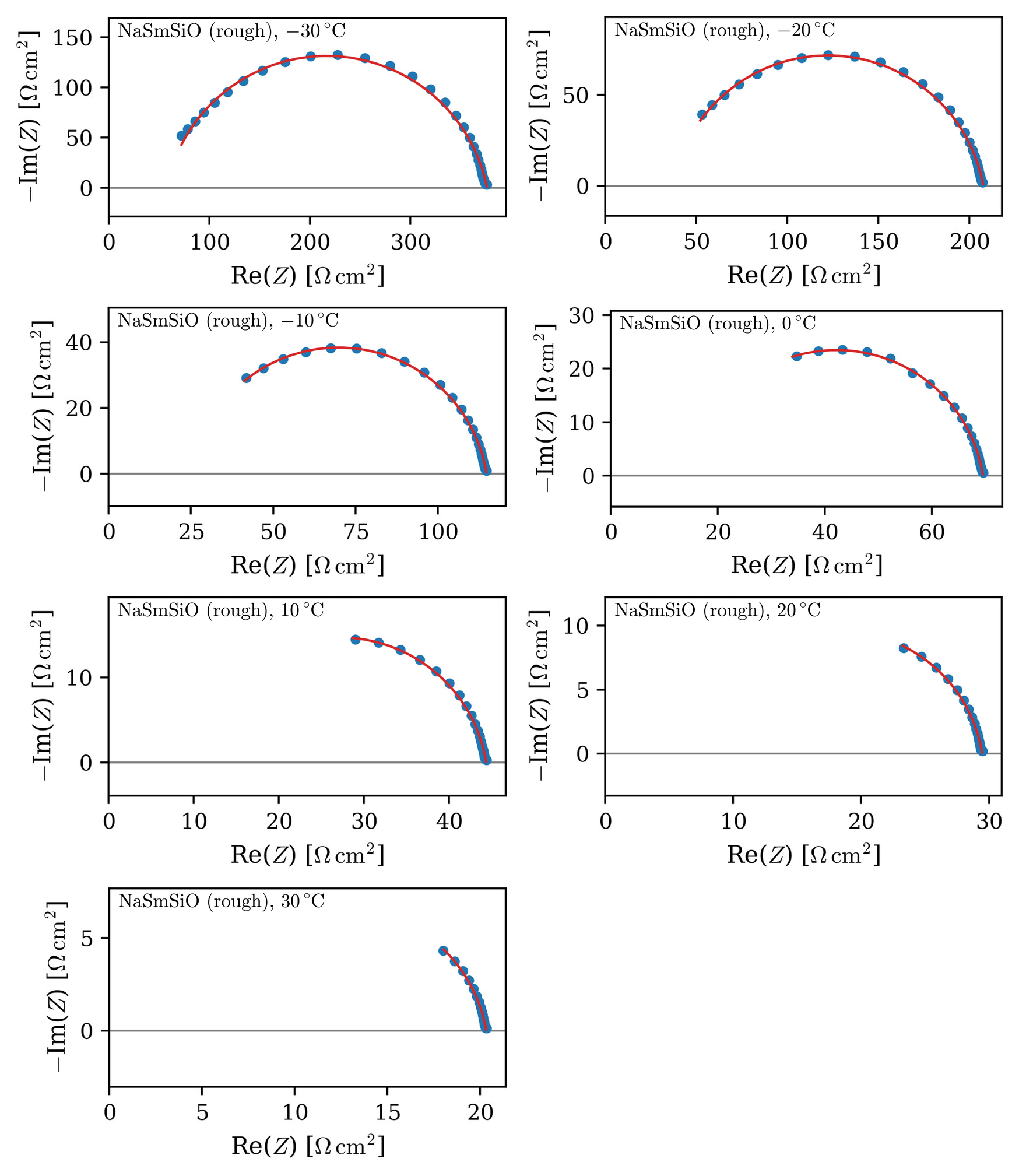}
\caption{Nyquist plots of the individual impedance spectra of symmetric
Na$|$\NaSmSiO$|$Na cells with a \emph{rough} electrolyte surface between $-30$ and
$+30\,^\circ\mathrm{C}$. Symbols: data; lines: fits to the equivalent circuit.}
\label{fig:nyq-nasmsio-rau}
\end{figure}

Figures~\ref{fig:nyq-nasmsio-pol} and~\ref{fig:nyq-nasmsio-rau} show every
individual spectrum behind Figs.~\ref{fig:impedance} and~\ref{fig:arrhenius} of the main text, for the polished and
the roughened electrolyte surface. Table~\ref{tab:fit-nasmsio} lists the complete
set of equivalent-circuit parameters.

The fitted interfacial element is $2\Rint$, because the two interfaces of the
symmetric cell are in series. The main text quotes $\Rint$ per interface, that
is, half of the tabulated value. A value of zero for $2\Rint$ means that the
interfacial arc was not separable and the parameter was fixed to zero, so that the
whole spectrum is attributed to $\RSE=\Rg+\Rgb$ (Section~\ref{sec:res-int} of the main text).

\begin{sidewaystable}
  \centering
  \footnotesize
  \setlength{\tabcolsep}{2pt}
  \caption{Equivalent-circuit parameters of symmetric Na$|$\NaSmSiO$|$Na cells
  between $-30$ and $+30\,^\circ\mathrm{C}$; top: polished, bottom: rough
  electrolyte surface. Values are fit value $\pm$ standard error of the fit.
  Resistances are area-specific (polished: $A=0.50\,\mathrm{cm^2}$, rough:
  $A=1.13\,\mathrm{cm^2}$). $\Rg$ and $\Rgb$ describe the grain and grain-boundary
  process of the electrolyte, and $2\Rint$ the two interfaces in series. $K$ and
  $\alpha$ are the constant-phase-element parameters. The derived time constants
  are $\tau=(RK)^{1/\alpha}$, evaluated from the \emph{absolute} resistance $R/A$
  rather than from the area-specific value tabulated here. ``--'' marks parameters
  that could not be determined meaningfully, and $2\Rint=0$ means the parameter
  was fixed to zero.}
  \label{tab:fit-nasmsio}
  \textbf{Polished surface}\\[3pt]
  \begin{tabular}{cccccccccc}
    \hline
    \fithead
    \hline
    $-30$ & $51.5 \pm 7.5$ & $(2.05 \pm 0.42) \cdot 10^{2}$ & $(4.08 \pm 0.40) \cdot 10^{2}$ & $(2.39 \pm 2.05) \cdot 10^{-8}$ & $(1.15 \pm 0.18) \cdot 10^{-6}$ & $0.868 \pm 0.072$ & $0.684 \pm 0.030$ & $(1.68 \pm 2.53) \cdot 10^{-6}$ & $(3.77 \pm 1.97) \cdot 10^{-5}$ \\
    $-20$ & $42.8 \pm 5.3$ & $(1.60 \pm 0.21) \cdot 10^{2}$ & $(3.03 \pm 0.19) \cdot 10^{2}$ & $(2.09 \pm 1.42) \cdot 10^{-8}$ & $(9.23 \pm 1.26) \cdot 10^{-7}$ & $0.878 \pm 0.055$ & $0.725 \pm 0.023$ & $(1.28 \pm 1.49) \cdot 10^{-6}$ & $(3.28 \pm 1.26) \cdot 10^{-5}$ \\
    $-10$ & $30.0 \pm 3.6$ & $87.0 \pm 10.5$ & $(1.68 \pm 0.09) \cdot 10^{2}$ & $(1.34 \pm 0.93) \cdot 10^{-8}$ & $(9.17 \pm 1.22) \cdot 10^{-7}$ & $0.919 \pm 0.055$ & $0.744 \pm 0.020$ & $(7.43 \pm 8.53) \cdot 10^{-7}$ & $(1.91 \pm 0.66) \cdot 10^{-5}$ \\
    $0$ & $21.5 \pm 2.2$ & $52.6 \pm 4.8$ & $94.1 \pm 3.4$ & $(1.13 \pm 0.65) \cdot 10^{-8}$ & $(7.41 \pm 0.80) \cdot 10^{-7}$ & $0.936 \pm 0.046$ & $0.783 \pm 0.014$ & $(4.67 \pm 4.42) \cdot 10^{-7}$ & $(1.20 \pm 0.30) \cdot 10^{-5}$ \\
    $10$ & $15.6 \pm 4.1$ & $27.2 \pm 7.0$ & $63.6 \pm 3.8$ & $(4.96 \pm 7.64) \cdot 10^{-9}$ & $(1.27 \pm 0.21) \cdot 10^{-6}$ & $1.00 \pm 0.12$ & $0.736 \pm 0.023$ & $(2.70 \pm 6.60) \cdot 10^{-7}$ & $(7.03 \pm 3.07) \cdot 10^{-6}$ \\
    $20$ & $12.4 \pm 4.4$ & $16.6 \pm 6.2$ & $40.6 \pm 2.4$ & $(5.37 \pm 10.63) \cdot 10^{-9}$ & $(1.36 \pm 0.20) \cdot 10^{-6}$ & $1.00 \pm 0.17$ & $0.744 \pm 0.020$ & $(1.78 \pm 5.82) \cdot 10^{-7}$ & $(4.81 \pm 1.86) \cdot 10^{-6}$ \\
    $30$ & $9.02 \pm 7.13$ & $11.9 \pm 8.5$ & $27.2 \pm 1.7$ & $(4.78 \pm 14.08) \cdot 10^{-9}$ & $(1.13 \pm 0.17) \cdot 10^{-6}$ & $1.00 \pm 0.27$ & $0.774 \pm 0.020$ & $(1.14 \pm 5.96) \cdot 10^{-7}$ & $(3.64 \pm 1.40) \cdot 10^{-6}$ \\
    \hline
  \end{tabular}
  \\[14pt]
  \textbf{Rough surface}\\[3pt]
  \begin{tabular}{cccccccccc}
    \hline
    \fithead
    \hline
    $-30$ & $57.7 \pm 1.3$ & $(3.19 \pm 0.02) \cdot 10^{2}$ & $0$ & $(4.80 \pm 0.24) \cdot 10^{-8}$ & -- & $0.877 \pm 0.004$ & -- & $(2.81 \pm 0.24) \cdot 10^{-6}$ & -- \\
    $-20$ & $37.1 \pm 0.7$ & $(1.71 \pm 0.01) \cdot 10^{2}$ & $0$ & $(4.63 \pm 0.20) \cdot 10^{-8}$ & -- & $0.888 \pm 0.004$ & -- & $(1.56 \pm 0.12) \cdot 10^{-6}$ & -- \\
    $-10$ & $25.0 \pm 0.3$ & $89.8 \pm 0.4$ & $0$ & $(4.36 \pm 0.14) \cdot 10^{-8}$ & -- & $0.899 \pm 0.003$ & -- & $(8.49 \pm 0.47) \cdot 10^{-7}$ & -- \\
    $0$ & $14.8 \pm 0.4$ & $54.8 \pm 0.4$ & $0$ & $(4.11 \pm 0.17) \cdot 10^{-8}$ & -- & $0.901 \pm 0.004$ & -- & $(4.73 \pm 0.37) \cdot 10^{-7}$ & -- \\
    $10$ & $9.99 \pm 0.39$ & $34.4 \pm 0.4$ & $0$ & $(4.18 \pm 0.19) \cdot 10^{-8}$ & -- & $0.901 \pm 0.004$ & -- & $(2.84 \pm 0.26) \cdot 10^{-7}$ & -- \\
    $20$ & $7.01 \pm 0.45$ & $22.5 \pm 0.5$ & $0$ & $(4.13 \pm 0.19) \cdot 10^{-8}$ & -- & $0.899 \pm 0.005$ & -- & $(1.70 \pm 0.19) \cdot 10^{-7}$ & -- \\
    $30$ & $4.66 \pm 0.65$ & $15.7 \pm 0.7$ & $0$ & $(3.98 \pm 0.20) \cdot 10^{-8}$ & -- & $0.893 \pm 0.007$ & -- & $(9.80 \pm 1.47) \cdot 10^{-8}$ & -- \\
    \hline
  \end{tabular}
\end{sidewaystable}

\clearpage

\section{Effect of humid-air exposure}
\label{sec:humidity}

\begin{figure}[htb]
\centering
\includegraphics[width=0.9\linewidth]{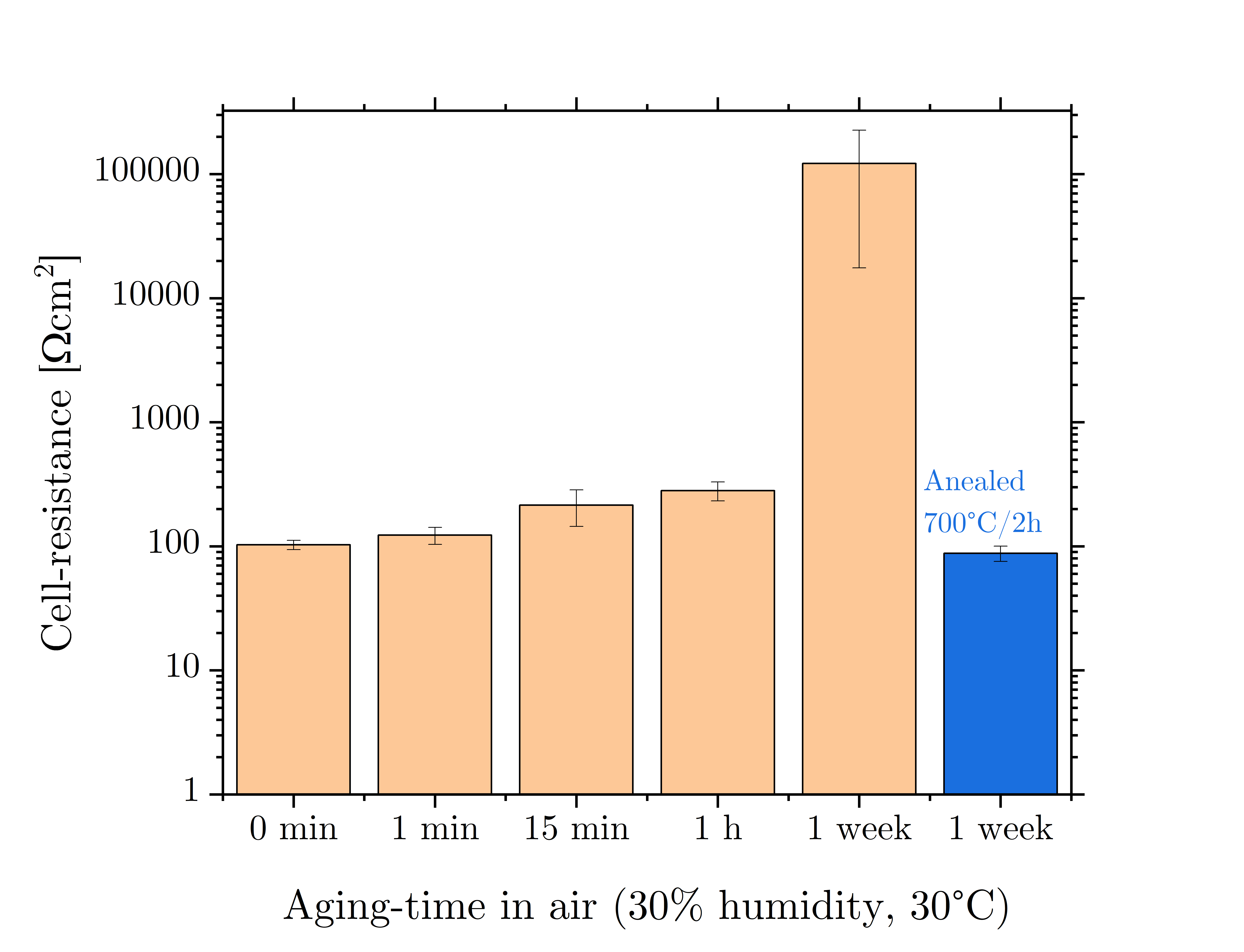}
\caption{Cell resistance of symmetric Na$|$\NaSmSiO$|$Na cells at
$30\,^\circ\mathrm{C}$ after the electrolyte had been exposed to air at
$\sim30\,\%$ relative humidity for the indicated time, with and without a
subsequent anneal at $700\,^\circ\mathrm{C}$ in argon.}
\label{fig:humidity}
\end{figure}

NaRSiO phases decompose on contact with moisture. Fig.~\ref{fig:humidity}
documents the cell resistance of symmetric Na$|$\NaSmSiO$|$Na cells whose
electrolyte had been exposed to air at $\sim30\,\%$ relative humidity before the
sodium electrodes were applied under argon. Annealing the exposed electrolyte at
$700\,^\circ\mathrm{C}$ in argon before contacting removes the degradation layer
again. This is the experiment referred to in Section~\ref{sec:exp} of the main text and the
reason why all samples reported there were ground and contacted exclusively under
argon.

\section{Validation of the method on Na\texorpdfstring{$|$}{|}NaSICON\texorpdfstring{$|$}{|}Na}
\label{sec:nasicon}

The analysis used in the main text rests on two steps that are worth testing
against an independent result: the attribution of the mid-frequency arc to the
interface, and the comparison of $\EA(\Rint)$ with $\EA(\RSE)$ as a criterion for
a contact-limited interface. NaSICON is the natural reference, because it is the
only sodium/oxide interface for which a well-characterised literature value
exists. Ortmann et al.\ report $\Rint\approx2\ohmcm$. Using high-resolution
impedance spectroscopy and degradation measurements, they show that this is the
limit at high contact pressure, that is, the value without a significant
constriction contribution. They also show that it originates from a kinetically
self-passivating decomposition layer \citep{Ortmann.2023}.

The measurement reported here differs from theirs in both preparation and
evaluation. The sodium foil is pressed onto the ceramic at $97\,^\circ\mathrm{C}$
and $10\,\mathrm{MPa}$. The cell is then held under load at
$30\,^\circ\mathrm{C}$ for at least two hours before the first spectrum, so the
self-passivating interphase has already formed when the measurement starts.
The interfacial contribution is separated by its temperature dependence
between $-30$ and $30\,^\circ\mathrm{C}$, rather than by a pressure series or by
extending the frequency window. An agreement between the two is therefore a
cross-check of two independent routes to the same quantity.

\begin{figure}[htb]
\centering
\includegraphics[width=0.78\linewidth]{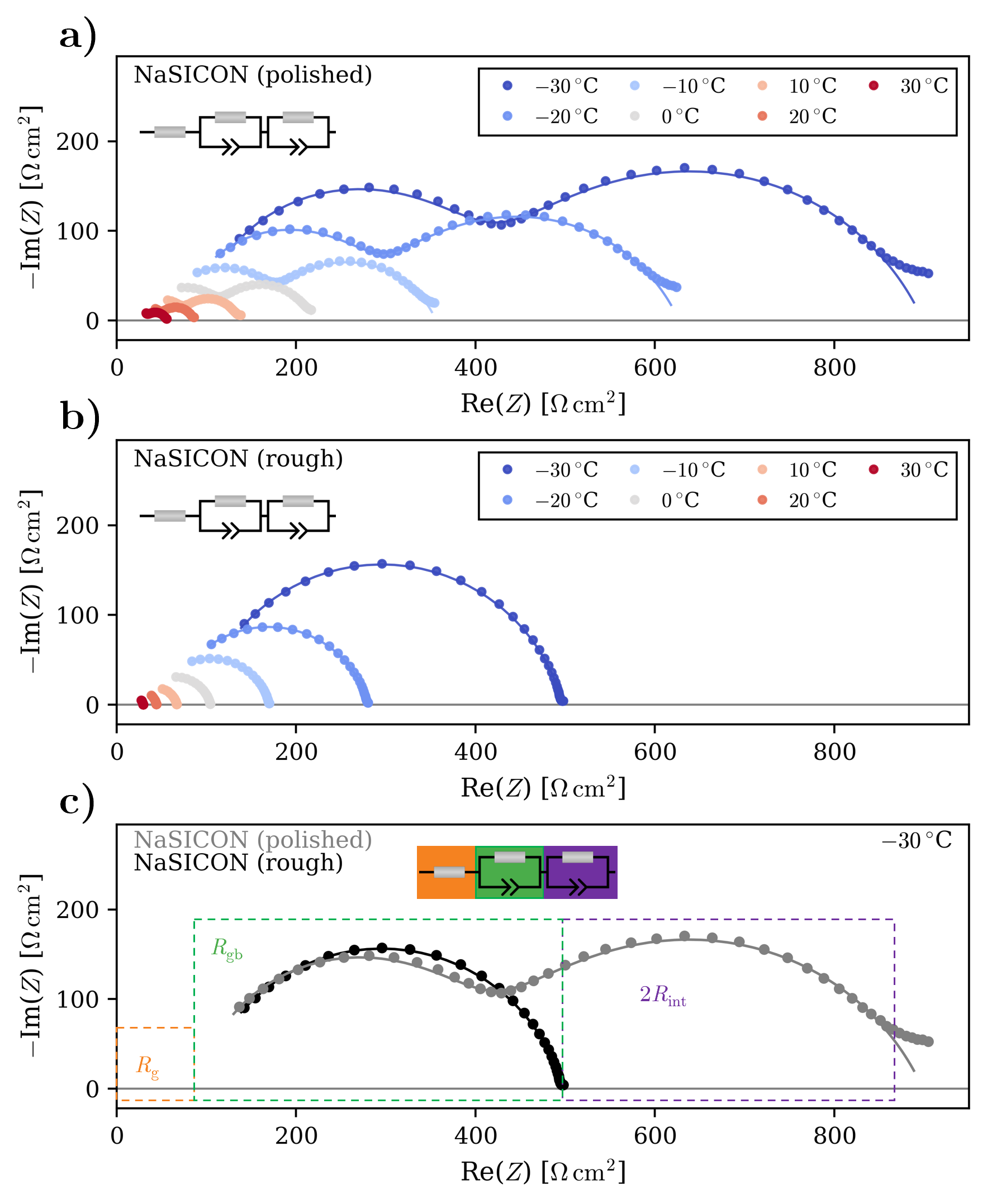}
\caption{Nyquist plots of symmetric Na$|$NaSICON$|$Na cells: (a) polished and
(b) rough electrolyte surfaces between $-30$ and $30\,^\circ\mathrm{C}$, and
(c) the two $-30\,^\circ\mathrm{C}$ spectra superimposed with the assignment of
the equivalent-circuit elements. As for \NaSmSiO, roughening removes the
interfacial arc. Symbols: data; lines: fits.}
\label{fig:si-imp}
\end{figure}

\begin{figure}[htb]
\centering
\includegraphics[width=0.80\linewidth]{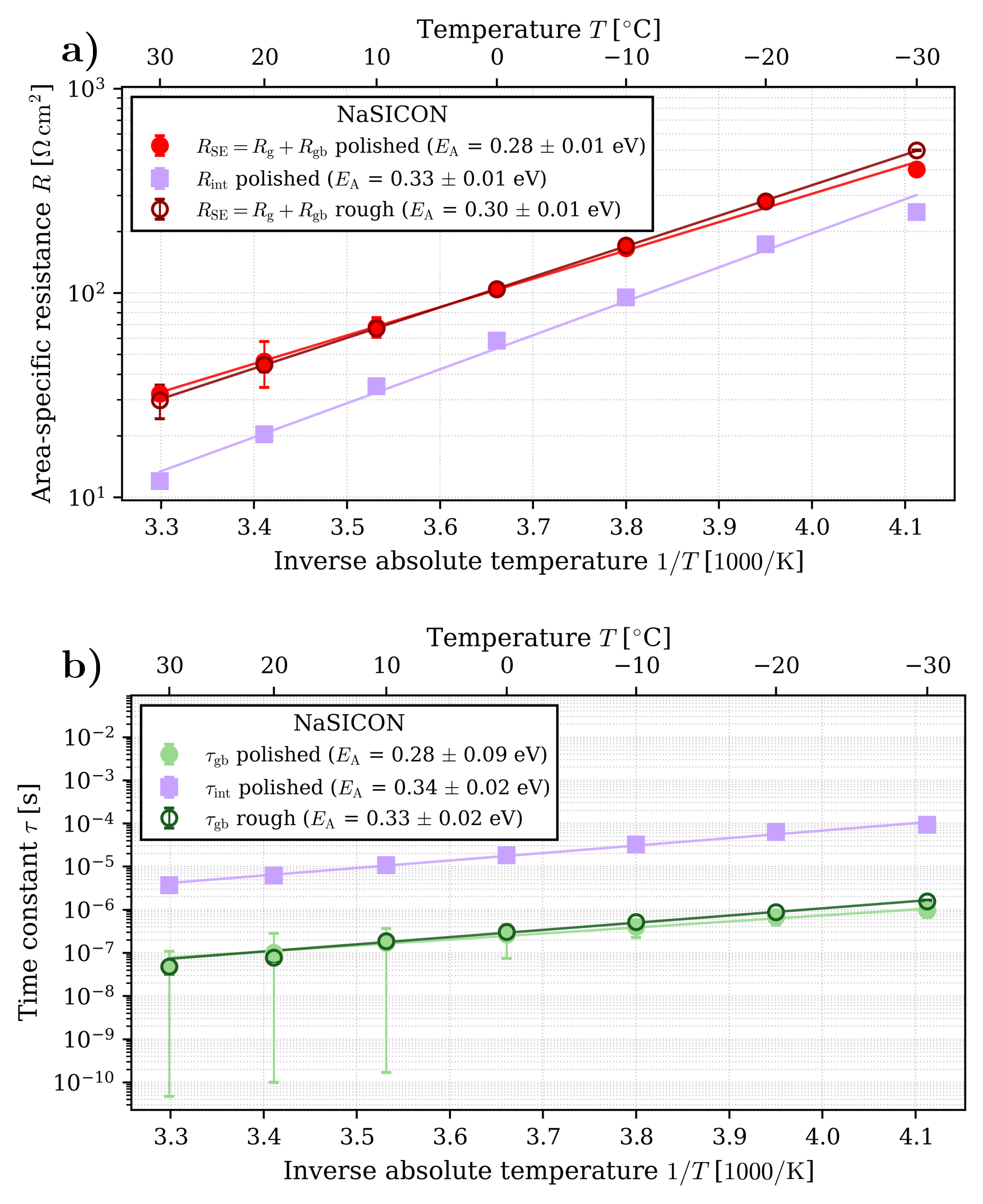}
\caption{Arrhenius plots for Na$|$NaSICON$|$Na between $-30$ and
$30\,^\circ\mathrm{C}$: (a) effective electrolyte resistance $\RSE$ for both
surface finishes and interfacial resistance $\Rint$ of the polished cell;
(b) the corresponding time constants $\taugb$ and $\tauint$.}
\label{fig:si-arr}
\end{figure}

\begin{figure}[htb]
\centering
\includegraphics[width=\linewidth]{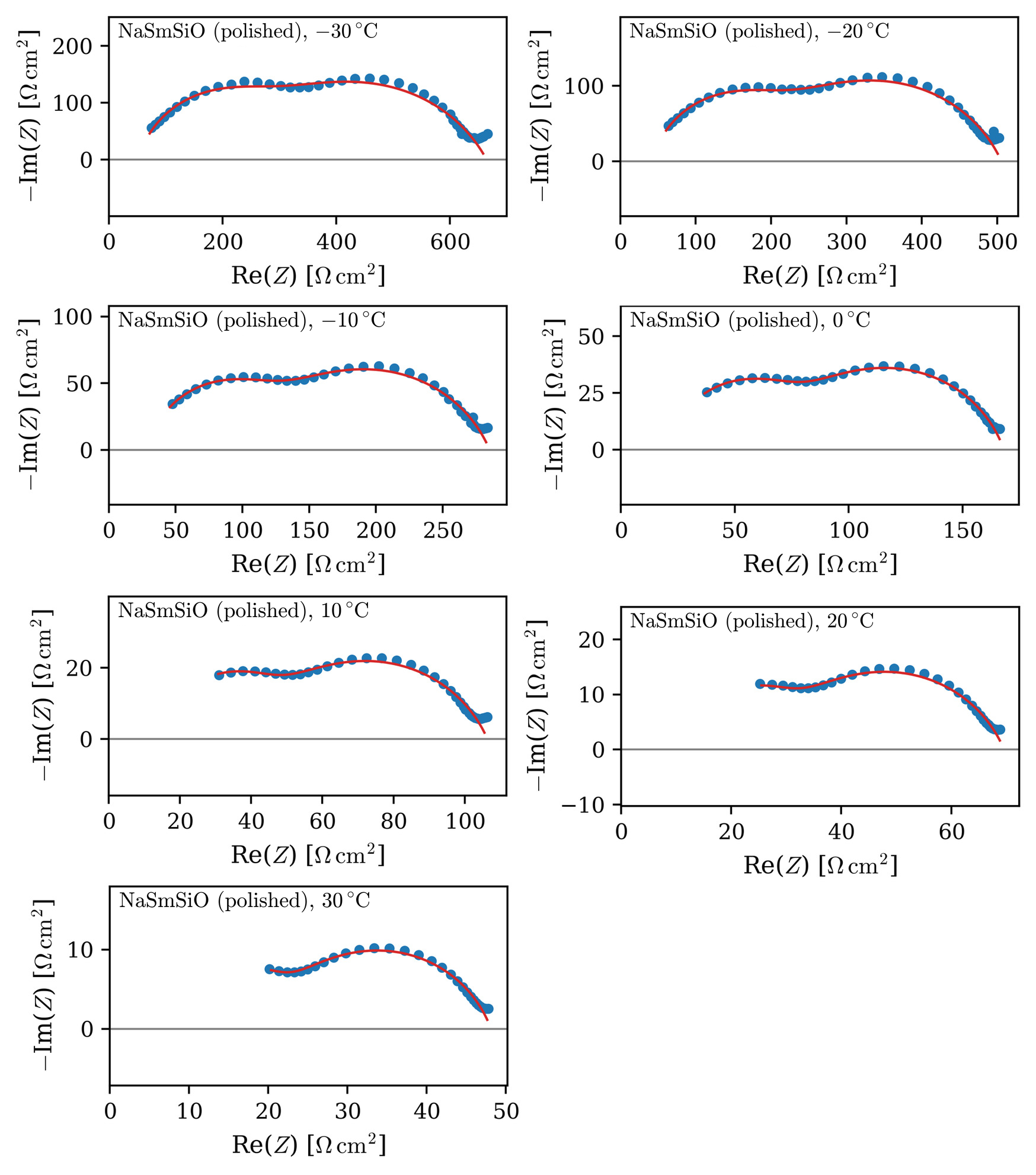}
\caption{Nyquist plots of the individual impedance spectra of symmetric
Na$|$NaSICON$|$Na cells with a \emph{polished} electrolyte surface between $-30$
and $+30\,^\circ\mathrm{C}$. Symbols: data; lines: fits.}
\label{fig:nyq-nasicon-pol}
\end{figure}

\begin{figure}[htb]
\centering
\includegraphics[width=\linewidth]{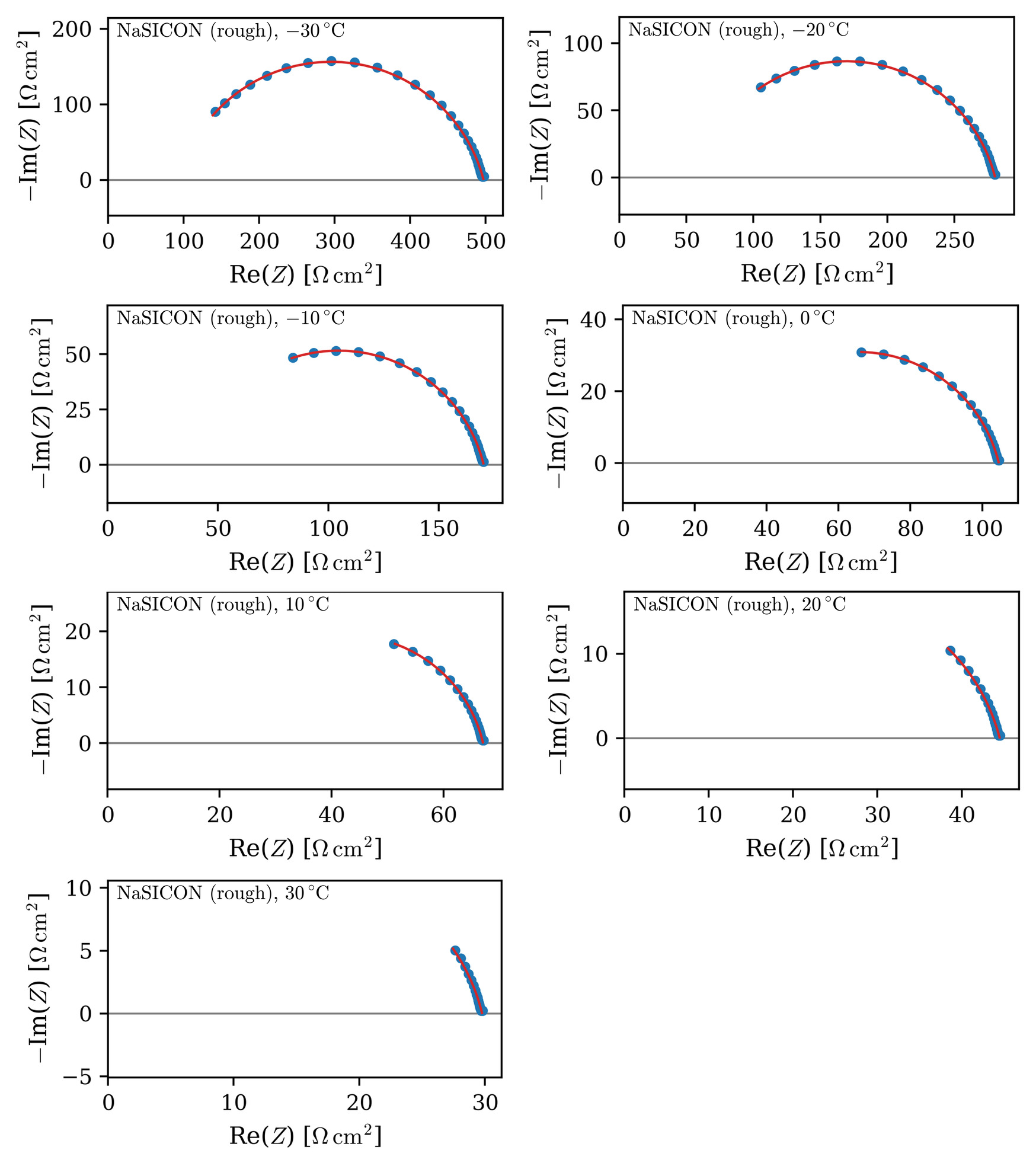}
\caption{Nyquist plots of the individual impedance spectra of symmetric
Na$|$NaSICON$|$Na cells with a \emph{rough} electrolyte surface between $-30$ and
$+30\,^\circ\mathrm{C}$. Symbols: data; lines: fits.}
\label{fig:nyq-nasicon-rau}
\end{figure}

The NaSICON pellets (\NaSICONc, solution-assisted solid-state reaction
\citep{Ma.2016, Ma.2019b}, sintered at $1350\,^\circ\mathrm{C}$ for
$30\,\mathrm{min}$, $\varnothing\,12\,\mathrm{mm}\times1.3\,\mathrm{mm}$,
relative density $>95\,\%$, $\sigma(30\,^\circ\mathrm{C})\approx
2\,\mathrm{mS\,cm^{-1}}$) were prepared, contacted and measured exactly as the
\NaSmSiO\ cells of the main text, with a polished (P4000) and a roughened (P360)
surface. Figures~\ref{fig:si-imp} and~\ref{fig:si-arr} summarise the spectra and
the Arrhenius analysis, Figs.~\ref{fig:nyq-nasicon-pol}
and~\ref{fig:nyq-nasicon-rau} show the individual spectra, and
Table~\ref{tab:fit-nasicon} lists the fit parameters.

\textbf{Polished surface.} Two polarisation processes are resolved
(Fig.~\ref{fig:si-imp}a,c), separated by about two orders of magnitude in time
constant. $\tauint$ runs from $(9.21\pm1.30)\cdot10^{-5}\,\mathrm{s}$ to
$(3.70\pm0.95)\cdot10^{-6}\,\mathrm{s}$ and $\taugb$ from
$(9.74\pm3.17)\cdot10^{-7}\,\mathrm{s}$ to $(4.76\pm6.14)\cdot10^{-8}\,\mathrm{s}$
(Table~\ref{tab:fit-nasicon}). The interfacial resistance is of the order of the
electrolyte resistance ($\Rint=12.0\pm0.4\ohmcm$ at $30\,^\circ\mathrm{C}$,
against $\RSE=32.1\pm0.6\ohmcm$). Its activation energy,
$\EA(\Rint)=0.33\pm0.01\,\mathrm{eV}$, exceeds that of the electrolyte,
$\EA(\RSE)=0.28\pm0.01\,\mathrm{eV}$, by $0.05\,\mathrm{eV}$. That is small on the
scale expected for a barrier of different physical origin, so the constriction
resistance is identified as the dominant contribution, exactly as for \NaSmSiO.

\textbf{Rough surface.} A single arc remains even at $-30\,^\circ\mathrm{C}$
(Fig.~\ref{fig:si-imp}b). As in the main text, $2\Rint$ was fixed to zero and the
whole spectrum attributed to $\RSE$. The remaining arc keeps the grain-boundary
time constant of the polished cell, $(1.57\pm0.08)\cdot10^{-6}$ to
$(4.83\pm1.66)\cdot10^{-8}\,\mathrm{s}$. Its activation energy,
$\EA(\RSE)=0.30\pm0.01\,\mathrm{eV}$, agrees with the polished value. With the
same conservative criterion, $\Rint\lesssim\RSE/20$ and
$\RSE(30\,^\circ\mathrm{C})=29.8\pm4.0\ohmcm$, this gives
\begin{equation*}
  \Rint \lesssim 1.5\ohmcm \quad (30\,^\circ\mathrm{C}),
\end{equation*}
below the $\approx2\ohmcm$ of Ortmann et al.\ and therefore compatible with it
\citep{Ortmann.2023}. Since the self-passivating interphase has formed before the
first spectrum is taken, that contribution is contained in the bound.

Two conclusions follow for the main text. First, the assignment of the
mid-frequency arc and the activation-energy criterion reproduce, on an
independent system and with a different experimental route, a literature value
obtained by a different method. Second, the qualitative behaviour of NaSICON and
\NaSmSiO\ is the same: constriction-dominated when polished, and no separable
interfacial arc when roughened. This supports treating the \NaSmSiO\ result as
a property of the preparation rather than of the silicate framework.

\begin{sidewaystable}
  \centering
  \footnotesize
  \setlength{\tabcolsep}{2pt}
  \caption{Equivalent-circuit parameters of symmetric Na$|$NaSICON$|$Na cells
  between $-30$ and $+30\,^\circ\mathrm{C}$; top: polished, bottom: rough
  electrolyte surface. Values are fit value $\pm$ standard error of the fit.
  Resistances are area-specific ($A=1.13\,\mathrm{cm^2}$). Notation and
  conventions as in Table~\ref{tab:fit-nasmsio}; $2\Rint$ is the fitted element
  for the two interfaces in series, so $\Rint$ per interface is half the
  tabulated value. $\Rg=0$ means that the fit no longer supports a separation of
  $\Rg$ from $\Rgb$ at that temperature, because the grain-boundary arc has moved
  too close to the high-frequency limit of the setup; the whole electrolyte
  contribution is then carried by $\Rgb$, and $\RSE=\Rg+\Rgb$ remains
  well determined.}
  \label{tab:fit-nasicon}
  \textbf{Polished surface}\\[3pt]
  \begin{tabular}{cccccccccc}
    \hline
    \fithead
    \hline
    $-30$ & $89.7 \pm 4.0$ & $(3.12 \pm 0.08) \cdot 10^{2}$ & $(4.98 \pm 0.07) \cdot 10^{2}$ & $(1.76 \pm 0.34) \cdot 10^{-8}$ & $(2.58 \pm 0.16) \cdot 10^{-6}$ & $0.884 \pm 0.015$ & $0.730 \pm 0.009$ & $(9.74 \pm 3.17) \cdot 10^{-7}$ & $(9.21 \pm 1.30) \cdot 10^{-5}$ \\
    $-20$ & $70.5 \pm 3.3$ & $(2.10 \pm 0.06) \cdot 10^{2}$ & $(3.46 \pm 0.05) \cdot 10^{2}$ & $(1.59 \pm 0.33) \cdot 10^{-8}$ & $(2.76 \pm 0.17) \cdot 10^{-6}$ & $0.896 \pm 0.017$ & $0.731 \pm 0.008$ & $(6.79 \pm 2.41) \cdot 10^{-7}$ & $(6.30 \pm 0.88) \cdot 10^{-5}$ \\
    $-10$ & $48.7 \pm 2.9$ & $(1.17 \pm 0.05) \cdot 10^{2}$ & $(1.91 \pm 0.03) \cdot 10^{2}$ & $(1.14 \pm 0.30) \cdot 10^{-8}$ & $(2.75 \pm 0.18) \cdot 10^{-6}$ & $0.927 \pm 0.021$ & $0.742 \pm 0.008$ & $(4.03 \pm 1.79) \cdot 10^{-7}$ & $(3.22 \pm 0.47) \cdot 10^{-5}$ \\
    $0$ & $37.9 \pm 3.6$ & $66.3 \pm 4.8$ & $(1.17 \pm 0.02) \cdot 10^{2}$ & $(5.59 \pm 2.47) \cdot 10^{-9}$ & $(3.12 \pm 0.26) \cdot 10^{-6}$ & $0.986 \pm 0.037$ & $0.736 \pm 0.010$ & $(2.67 \pm 1.92) \cdot 10^{-7}$ & $(1.81 \pm 0.34) \cdot 10^{-5}$ \\
    $10$ & $26.5 \pm 4.9$ & $41.7 \pm 5.8$ & $69.9 \pm 1.5$ & $(4.60 \pm 3.11) \cdot 10^{-9}$ & $(3.26 \pm 0.30) \cdot 10^{-6}$ & $1.00 \pm 0.06$ & $0.744 \pm 0.011$ & $(1.70 \pm 1.96) \cdot 10^{-7}$ & $(1.07 \pm 0.23) \cdot 10^{-5}$ \\
    $20$ & $17.0 \pm 7.9$ & $29.2 \pm 8.6$ & $40.7 \pm 1.1$ & $(3.87 \pm 3.67) \cdot 10^{-9}$ & $(2.99 \pm 0.29) \cdot 10^{-6}$ & $1.00 \pm 0.09$ & $0.761 \pm 0.012$ & $(9.99 \pm 18.19) \cdot 10^{-8}$ & $(6.10 \pm 1.39) \cdot 10^{-6}$ \\
    $30$ & $0$ & $32.1 \pm 0.6$ & $24.0 \pm 0.7$ & $(1.67 \pm 1.46) \cdot 10^{-9}$ & $(3.08 \pm 0.35) \cdot 10^{-6}$ & $1.00 \pm 0.06$ & $0.770 \pm 0.013$ & $(4.76 \pm 6.14) \cdot 10^{-8}$ & $(3.70 \pm 0.95) \cdot 10^{-6}$ \\
    \hline
  \end{tabular}
  \\[14pt]
  \textbf{Rough surface}\\[3pt]
  \begin{tabular}{cccccccccc}
    \hline
    \fithead
    \hline
    $-30$ & $93.6 \pm 1.2$ & $(4.04 \pm 0.01) \cdot 10^{2}$ & $0$ & $(3.84 \pm 0.11) \cdot 10^{-8}$ & -- & $0.838 \pm 0.002$ & -- & $(1.57 \pm 0.08) \cdot 10^{-6}$ & -- \\
    $-20$ & $58.7 \pm 0.6$ & $(2.22 \pm 0.01) \cdot 10^{2}$ & $0$ & $(3.94 \pm 0.09) \cdot 10^{-8}$ & -- & $0.844 \pm 0.002$ & -- & $(8.74 \pm 0.37) \cdot 10^{-7}$ & -- \\
    $-10$ & $39.0 \pm 0.5$ & $(1.31 \pm 0.01) \cdot 10^{2}$ & $0$ & $(4.07 \pm 0.09) \cdot 10^{-8}$ & -- & $0.847 \pm 0.002$ & -- & $(5.19 \pm 0.22) \cdot 10^{-7}$ & -- \\
    $0$ & $26.4 \pm 0.5$ & $78.3 \pm 0.5$ & $0$ & $(4.15 \pm 0.11) \cdot 10^{-8}$ & -- & $0.851 \pm 0.003$ & -- & $(3.07 \pm 0.17) \cdot 10^{-7}$ & -- \\
    $10$ & $18.7 \pm 0.7$ & $48.4 \pm 0.7$ & $0$ & $(4.21 \pm 0.15) \cdot 10^{-8}$ & -- & $0.854 \pm 0.004$ & -- & $(1.88 \pm 0.17) \cdot 10^{-7}$ & -- \\
    $20$ & $0$ & $44.5 \pm 2.5$ & $0$ & $(2.78 \pm 0.15) \cdot 10^{-8}$ & -- & $0.838 \pm 0.008$ & -- & $(7.80 \pm 1.47) \cdot 10^{-8}$ & -- \\
    $30$ & $0$ & $29.8 \pm 4.0$ & $0$ & $(2.35 \pm 0.31) \cdot 10^{-8}$ & -- & $0.848 \pm 0.013$ & -- & $(4.83 \pm 1.66) \cdot 10^{-8}$ & -- \\
    \hline
  \end{tabular}
\end{sidewaystable}

\clearpage
\section{Thermodynamic stability of NaSICON against liquid sodium}
\label{sec:nasicon-immersion}

The immersion experiment of Section~\ref{sec:res-stab} of the main text was repeated on NaSICON
under identical conditions, so that the behaviour of the two oxide families can be
compared on samples treated the same way rather than across the literature.

\begin{figure}[htb]
\centering
\includegraphics[width=0.56\linewidth]{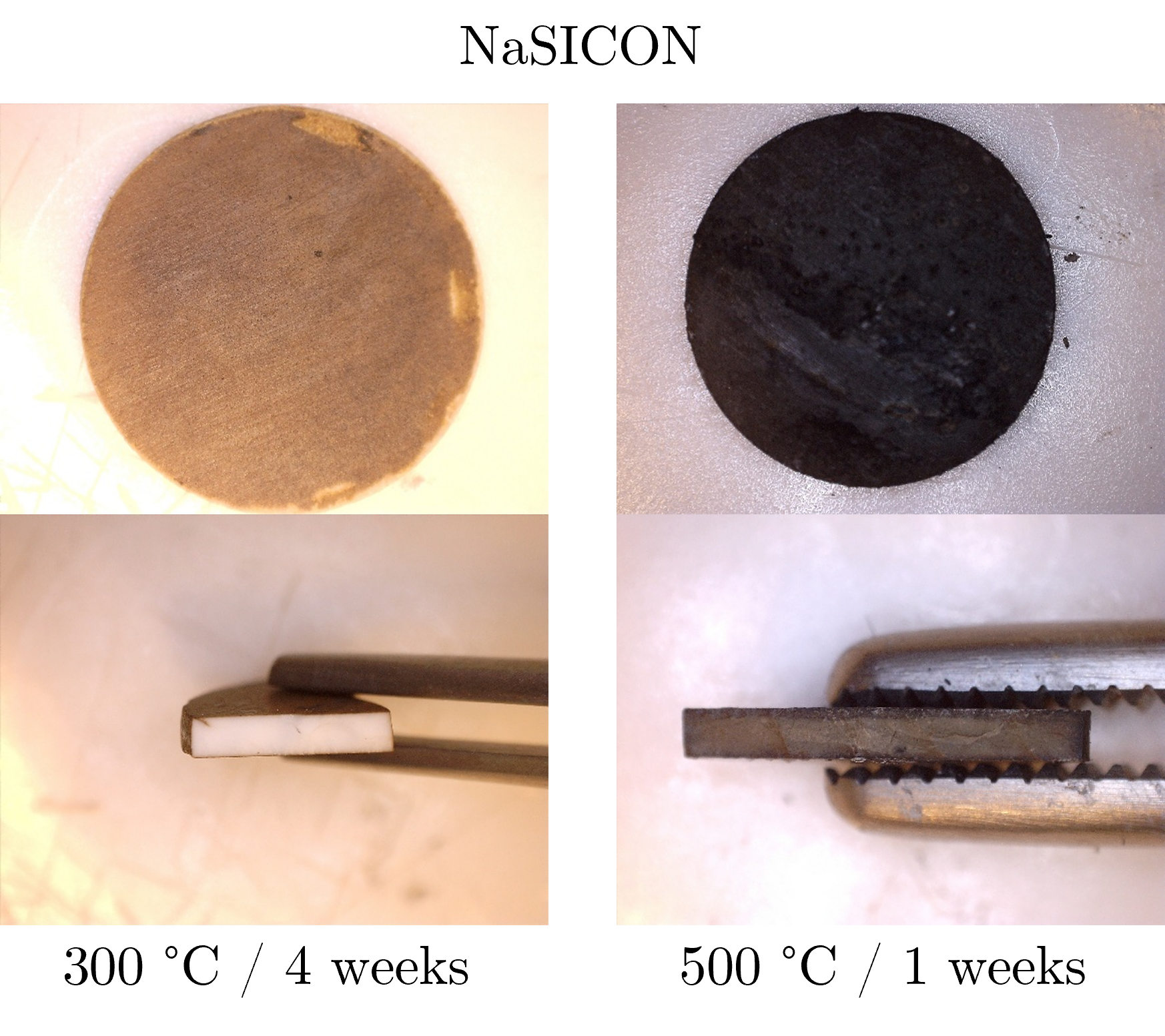}
\caption{NaSICON pellets after immersion in liquid sodium. Left: after four weeks
at $300\,^\circ\mathrm{C}$. The surface is discoloured (top) while the fracture
cross-section remains white (bottom), i.e.\ the reaction front passivates. Right: a
second pellet after one week at $500\,^\circ\mathrm{C}$, black at the surface
(top) and through the entire cross-section (bottom). Pellet thickness of $\sim 1.3\,\mathrm{mm}$.}
\label{fig:nasicon-immersion}
\end{figure}

After four weeks in liquid sodium at $300\,^\circ\mathrm{C}$ the NaSICON pellet is
discoloured at the surface, but a fresh fracture cross-section is white throughout
(Fig.~\ref{fig:nasicon-immersion}, left). The reaction is therefore confined to a
surface layer: it passivates. A second pellet held in sodium melt at
$500\,^\circ\mathrm{C}$ for one week is black over the whole cross-section
(Fig.~\ref{fig:nasicon-immersion}, right), so the reaction is not absent at
$300\,^\circ\mathrm{C}$ but merely slow and self-limiting.

NaSICON is thus thermodynamically unstable against sodium metal, in agreement with
the literature \citep{schmid1982, Ortmann.2023}, while remaining usable because the
decomposition products conduct electrons poorly and the layer stops growing. This
is the behaviour \NaSmSiO\ and \NaYSiO\ do \emph{not} show. Under the same
conditions, $300\,^\circ\mathrm{C}$ and four weeks, their reaction front crosses
the entire pellet (Fig.~\ref{fig:immersion} of the main text). The distinction drawn in the
main text between a passivating and a non-passivating reaction front therefore
rests on a direct side-by-side comparison, not only on literature values. For
room-temperature operation the conclusion is the same for both families:
thermodynamically unstable, but with decomposition kinetics slow enough to be
irrelevant on the timescale of a cell.

\end{document}